\documentclass[11pt,a4paper]{article}

\usepackage[T1]{fontenc}
\usepackage[latin9]{inputenc}
\usepackage[a4paper]{geometry}
\usepackage{amsmath}
\usepackage{amsfonts}
\usepackage{amssymb} 
\usepackage{graphicx}
\usepackage{a4wide}
\usepackage{hepunits}
\usepackage{url}
\usepackage{supertabular}
\usepackage{xspace}
\usepackage{latexsym}
\usepackage{textcomp}
\usepackage{wrapfig}
\usepackage{subfig}
\usepackage{slashed}
\usepackage{setspace}
\usepackage[dvipsnames]{xcolor}
\usepackage{citesort}

\author{C.~Mezrag$^1$\footnote{cedric.mezrag@cea.fr}, H.~Moutarde$^1$\footnote{herve.moutarde@cea.fr},
 J.~Rodr\'iguez-Quintero$^2$\footnote{jose.rodriguez@dfaie.uhu.es}, F.~Sabati\'e$^1$\footnote{franck.sabatie@cea.fr} \\
{\small \textit{$^1$ CEA, Centre de Saclay, IRFU/Service de Physique Nucl\'eaire}} \\
{\small \textit{F-91191 Gif-sur-Yvette, France}} \\
{\small \textit{$^2$ Departamento de F\'isica Aplicada, Facultad de Ciencias Experimentales}} \\
{\small \textit{Universidad de Huelva, Huelva 21071, Spain.}}
}

\title {Towards a Pion Generalized Parton Distribution Model \\ From \dses}

\date{}

\DeclareMathOperator{\SymCovDev}{\overset{\leftrightarrow}D}

\newcommand{\refeq}[1]{Eq.~(\ref{#1})}
\newcommand{\refeqs}[2]{Eqs.~(\ref{#1}-\ref{#2})}

\newcommand{\reffig}[1]{Fig.~\ref{#1}}
\newcommand{\refcite}[1]{Ref.~\cite{#1}}
\newcommand{\refsec}[1]{Sec.~\ref{#1}}
\newcommand{\refapp}[1]{App.~\ref{#1}}

\newcommand{\ie}{\textit{i.e. }}
\newcommand{\eg}{\textit{e.g. }}
\newcommand{\etal}{\textit{et al.\xspace}}
\newcommand{\lhs}{l.h.s.\xspace}
\newcommand{\rhs}{r.h.s.\xspace}
\newcommand{\ds}{Dyson-Schwinger\xspace}
\newcommand{\dse}{Dyson-Schwinger equation\xspace}
\newcommand{\dses}{Dyson-Schwinger equations\xspace}

\newcommand{\bs}{Bethe-Salpeter\xspace}
\newcommand{\bse}{Bethe-Salpeter equation\xspace}

\newcommand{\TwistTwoQuarkOp}[3]{\ensuremath{\bar{q}\gamma^{\{#1} i \SymCovDev\phantom{D}\!\!\!\!\!^{#2}\ldots i \SymCovDev\phantom{D}\!\!\!\!\!^{#3\}}q}\xspace}

\newcommand{\ket}[1]{\ensuremath{\left|#1\right\rangle}\xspace}
\newcommand{\bra}[1]{\ensuremath{\left\langle #1\right|}\xspace}
\newcommand{\elt}[3]{\ensuremath{\left\langle #1 \left| #2 \right| #3 \right\rangle}\xspace} 
\newcommand{\cond}[1]{\ensuremath{\left\langle #1 \right\rangle}\xspace} 
\newcommand{\Dd}[0]{\ensuremath{\frac{\Delta}{2}}}

\newcommand{\Pn}[0]{\ensuremath{P \cdot n}}
\newcommand{\g}[0]{\ensuremath{\gamma}}
\newcommand{\G}[0]{\ensuremath{\Gamma}}
\newcommand{\wl}[2]{\ensuremath{\left[#1;#2\right]}\xspace}

\begin{document}

\maketitle

\begin{abstract}

We compute the pion quark Generalized Parton Distribution $H^q$ and quark Double Distributions $F^q$ and $G^q$ in a coupled \bs and \ds approach in terms of quarks flavors or isospin states. We use simple analytic expressions inspired by the numerical resolution of \ds and \bs equations. We explicitly check the support and polynomiality properties, and the behavior under charge conjugation or time invariance of our model. We obtain analytic expressions for the pion Double Distributions and Generalized Parton Distribution at vanishing pion momentum transfer at a low scale. Our model compare very well to experimental pion form factor or Parton Distribution Function data. This paper is the first stage of a GPD-modeling program which will be pursued by incorporating more realistic solutions of the \bs and \ds equations.

\end{abstract}


\section*{Introduction}
Generalized Parton Distributions (GPDs) were introduced independently by M\"uller \etal \cite{Mueller:1998fv}, Ji \cite{Ji:1996nm} and Radyushkin \cite{Radyushkin:1997ki}. They are related to hadron form factors by sum rules, and contain the usual Parton Distribution Functions (PDFs) as a limiting case. But they not only generalize the classical objects describing the static or dynamical content of hadrons; they also provide unique information about the structure of hadrons, including 3D imaging of their partonic components and access to the quark orbital angular momentum. GPDs have been the object of an intense theoretical and experimental activity ever since (see the reviews 
\refcite{Ji:1998pc,Goeke:2001tz,Diehl:2003ny,Belitsky:2005qn,Boffi:2007yc,Guidal:2013rya} and references therein).

Most of the theoretical constraints on GPDs are automatically fulfilled by modeling Double Distributions (DDs) \cite{Mueller:1998fv,Radyushkin:1998es,Radyushkin:1998bz}, which are Radon transform of GPDs \cite{Teryaev:2001qm}. DD modeling has been the most popular way to build realistic models from the early days of GPDs (see \eg the review \refcite{Guidal:2013rya} and references therein). Yet, these classical models, or alternative models like \eg in \refcite{Broniowski:2007si,Frederico:2009fk,Goldstein:2010gu,Mezrag:2013mya}, need at some points phenomenological parameterizations. Their comparison to experimental data meet some successes usually depending on the kinematic region. It is nevertheless neither clear how to improve them in a systematic way, nor how to achieve real predictive power for the GPDs that decouple in the forward limit. To obtain a better agreement with the data one may either develop more sophisticated GPD parameterizations, as advocated in \refcite{Kumericki:2008di}, or try different implementations of DD modeling. Elaborating on ideas employed since the nineties mostly in spectroscopy, some modeling tools have received considerable attention in recent years (see \eg the reviews \refcite{Roberts:1994dr,Alkofer:2000wg,Maris:2003vk,Bashir:2012fs,Roberts:2012sv}). They consist in implementing the \dses \cite{Dyson:1949ha,Schwinger:1951ex,Schwinger:1951hq} to describe the partonic dynamics in a hadron, and the \bse \cite{Schwinger:1951ex,Schwinger:1951hq,Salpeter:1951sz,GellMann:1951rw,Schwinger:1953tb} to switch from a partonic to an hadronic picture. We apply this strategy in the present paper. 

Most of the experimental attention has been devoted so far to nucleon GPDs, with Deeply Virtual Compton Scattering (DVCS) early recognized as a key channel to access GPDs, in particular in the valence region \cite{Guidal:2013rya}. Pion GPDs offer the theoretical advantage of a much simpler spin structure than nucleon GPDs, and are particularly interesting due to the special role of the pion with respect to chiral symmetry. Accessing pion GPDs from experiment is harder than nucleon GPDs although it seems feasible after the upgrade of Jefferson Lab at 12~\GeV\xspace  through the study of DVCS on a virtual pion target \cite{Amrath:2008vx}.

On the theoretical side, pion GPDs have been studied using different tools. Polyakov and Weiss \cite{Polyakov:1999gs}, and Anikin \etal \cite{Anikin:1999pf}, discussed the effect of an instanton vacuum by means of a effective nonlocal quark-hadron lagrangian. Chiral symmetry is also central in the developments of Broniowski \etal \cite{Broniowski:2003rp,Broniowski:2007si} in the framework of the Nambu~-~Jona~-~Lasinio model (see the reviews \refcite{RuizArriola:2002wr,Christov:1995vm} and references therein). Choi \etal \cite{Choi:2001fc,Choi:2002ic}, then Mukherjee and Radyushkin \cite{Mukherjee:2002gb}, proposed light-frond calculations with gaussian or power-law wavefunctions in a triangle diagram approximation.  Later Ji, Mishchenko and Radyushkin \cite{Ji:2006ea} discussed the relation between an higher-Fock component of $q\bar{q}g$ type and the nonzero value of the GPD at $x = \xi$. GPD modeling in the \bs framework has enjoyed several studies \cite{Frederico:2009fk,Tiburzi:2002tq,Theussl:2002xp,Bissey:2002yr,Bissey:2003yr,VanDyck:2007jt}, usually with simple \bs vertices and with computations of triangle diagrams. Note that the authors of \refcite{Frederico:2009fk} also discussed two other GPD models, but at the very specific values $\xi = 0$ or 1. Amrath \etal \cite{Amrath:2008vx} modeled the GPD $H$ in the framework of the popular Radyushkin Double Distribution Ansatz \cite{Musatov:1999xp}. Pion GPD modeling for large, or moderately large, $t$, was investigated by Bakulev \etal \cite{Bakulev:2000eb}, Vogt \cite{Vogt:2001if} and Hoodboy \etal \cite{Hoodbhoy:2003uu}. At last, let us mention the computation of the generalized form factors in chiral perturbation theory at one-loop order by Diehl \etal \cite{Diehl:2005rn}. This study focuses on applications to lattice QCD and does not proceed further to a complete model of the pion GPDs.

We develop here an original way to model GPDs, based on \ds and \bs equations, and DDs. In the first section of this paper we remind the general GPD and DD formalisms, specify the relevant kinematics and outline the known theoretical constraints. In the second section we describe the computation of pion GPDs in a simplified \bs and \ds approach. In the third section we compare our model to existing experimental data to conclude our study in a fourth section.


\section{Generalized Parton Distributions: theoretical framework}
\label{sec:GPD}

In this section we introduce the appropriate definitions and deal with the general GPD properties which are relevant for our study.


\subsection{Definition and isospin properties}
\label{sec:GPD_Definition}

For any four-vector $v$ we note:
\begin{equation}
\label{eq:def-lc-coordinates}
v^{\pm} = \frac{1}{\sqrt{2}} ( v^0 \pm v^3 ) \quad \textrm{ and } \quad v = ( v^+, v_\perp, v^-).
\end{equation}
$u \cdot v$ denotes the scalar product of two four-vectors $u$ and $v$. $n$ and $\tilde{n}$ are the light-cone vectors such that $v^+ = v \cdot n$ and $v^- = v \cdot \tilde{n}$.

The GPD $H^q_\pi$, $q$ denoting the quark flavor, is introduced through the matrix element of \refeq{eq-def-GPD-H-spinless-target}:
\begin{equation}
\label{eq-def-GPD-H-spinless-target}
H^q_{\pi}( x, \xi, t ) =  \frac{1}{2} \int \frac{\mathrm{d}z^-}{2\pi} \, e^{i x P^+ z^-} \bra{\pi,P+\frac{\Delta}{2}} \bar{q}\left(-\frac{z}{2}\right)\gamma^+\wl{-\frac{z}{2}}{\frac{z}{2}}q\left(\frac{z}{2}\right) \ket{\pi,P-\frac{\Delta}{2}}_{z^+=0, z_\perp=0}, 
\end{equation}
where  the pion state can be either $\pi^-$, $\pi^0$ or $\pi^+$. We note $\xi = - \Delta^+/(2P^+)$ the skewness, $t=\Delta^2$ the momentum transfert, and $[ \cdot ; \cdot ]$ the Wilson line along a light-like path joining the two fields at position $-z/2$ and $+z/2$. Note that $P^2 = m_\pi^2 - t / 4$ where $m_\pi$ is the charged pion mass. In all the following we will adopt the light cone gauge, replacing everywhere the Wilson line by the identity.

In order to implement isospin symmetry in the system of pion GPDs, we note $\tau^1$, $\tau^2$ and $\tau^3$ the Pauli matrices, and we write $\tau^{\pm} = ( \tau^1 \pm i \tau^2 ) / \sqrt{2}$. The isosinglet and isovector GPDs $H^{I=0}$ and $H^{I=1}$ are defined in terms of the following matrix elements:
\begin{eqnarray}
  \label{eq:def_isoscalar}
\left\{ 
\begin{array}{c} 
 \delta^{ab} H^{I=0}(x,\xi,t) \\
   i\epsilon^{abc} H^{I=1}(x,\xi,t)
\end{array}
\right\}  & = & \frac{1}{2} \int \frac{\mathrm{d}z^-}{2\pi} \, e^{i x P^+ z^-} 
  \\ 
 & \times & 
 \bra{\pi^b,P+\frac{\Delta}{2}} \bar{\psi}\left(-\frac{z}{2}\right)
\left\{ \begin{array}{c}
\mathcal{I} \gamma^+ \\ 
\tau^c\gamma^+
\end{array} 
 \right\}
\wl{-\frac{z}{2}}{\frac{z}{2}}\psi\left(\frac{z}{2}\right) \ket{\pi^a,P-\frac{\Delta}{2}}_{z^+=0, z_\perp=0} \ , 
\nonumber
\end{eqnarray}
where $\psi$ denotes the doublet of $u$ and $d$ quark fields, $\mathcal{I}$ is the identity, $\tau^c$ the Pauli matrices, and $\ket{\pi^1}$, $\ket{\pi^2}$ and $\ket{\pi^3}$ is a cartesian basis of the adjoint representation of the Lie algebra $\mathfrak{su}(2)$. The vectors of this basis can be expressed in terms of charge eigenstates:
\begin{eqnarray}
  \label{eq:isospin_basis_charged_pions}
  \ket{\pi^\pm} & = & \frac{1}{\sqrt{2}}\left(\ket{\pi^1}\pm i\ket{\pi^2}\right), \\
 \label{eq:isospin_basis_neutral_pion}
 \ket{\pi^0} & = & \ket{\pi^3}.
\end{eqnarray}
Therefore 
we get:
\begin{eqnarray}
  H^{I=0}(x,\xi,t) 
& = & H^u_{\pi^\pm}(x,\xi,t) + H^d_{\pi^\pm}(x,\xi,t)   
\ = \ H^u_{\pi^0}(x,\xi,t) + H^d_{\pi^0}(x,\xi,t),   \label{eq:relation_isoscalar_neutral_pions} \\
 H^{I=1}(x,\xi,t) 
& = & H^u_{\pi^+}(x,\xi,t) - H^d_{\pi^+}(x,\xi,t)  
\ = \ - \big( H^u_{\pi^-}(x,\xi,t) - H^d_{\pi^-}(x,\xi,t)  \big), \label{eq:relation_isovector_piminus} \\
0
& = & H^u_{\pi^0}(x,\xi,t) - H^d_{\pi^0}(x,\xi,t).   \label{eq:relation_isovector_neutral_pions} 
\end{eqnarray}
From \refeq{eq:relation_isoscalar_neutral_pions} 
and \refeq{eq:relation_isovector_piminus} we deduce that:
\begin{eqnarray}
H^u_{\pi^{\pm}}(x,\xi,t) & = & H^d_{\pi^{\mp}}(x,\xi,t), 
\end{eqnarray}
and adding \refeq{eq:relation_isoscalar_neutral_pions} and \refeq{eq:relation_isovector_neutral_pions} we get:
\begin{equation}
H^u_{\pi^0}(x,\xi,t) = H^d_{\pi^0}(x,\xi,t) = \frac{1}{2} \big( H^u_{\pi^+}(x,\xi,t) + H^d_{\pi^+}(x,\xi,t) \big).
\end{equation}
Therefore isospin symmetry dictates that all the information for the whole system of pions GPDs (and thus of pion form factors or PDFs) can be equivalently encoded into $H^{I=0}$ and $H^{I=1}$ or $H^u_{\pi^+}$ and $H^d_{\pi^+}$. Unless explicitly stated otherwise, we will reserve the notations $H^u$ and $H^d$ for the $\pi^+$ state. The GPDs $H^u$ and $H^d$ are the objects we compute in \refsec{sec:GPD_modeling}.


\subsection{Properties from discrete symmetries and Lorentz invariance}
\label{subsec:Lorentz}

Discrete symmetries have interesting consequences on GPDs, which are most visible in the isospin representation. Charge conjugation, $H^q_{\pi^+}(x,\xi,t) =  - H^q_{\pi^-}(-x,\xi,t)$ for $q=u,d$, 
combined with isospin symmetry requirements further yields:
\begin{equation}
H^u_{\pi^+}(x,\xi,t) = - H^d_{\pi^+}(-x,\xi,t).
\label{eq:G-parity-GPD}
\end{equation}
In terms of the GPDs corresponding to the $\pi^+$ state, we get:
\begin{eqnarray}
H^I(x,\xi,t)
& = & H^u(x,\xi,t) + (-1)^{1-I} H^u(-x,\xi,t), \label{eq:H_isoscalar_both} 
\end{eqnarray}
which makes $H^{I=0}$ (resp. $H^{I=1}$) an odd (resp. even) function of $x$: 
$H^I(-x,\xi,t) = (-1)^{1-I} H^I(x,\xi,t)$ for $I = 0, 1$.

Analogous results hold in the forward limit ($\Delta=0$) 
and manifest also themselves in sum rules. From the expression of the electromagnetic current, we introduce the quark contributions $F^u_\pi( t )$ and $F^d_\pi( t )$ to the pion ($\pi^+$) form factor $F_\pi( t )$:
\begin{equation}
\label{eq:decomposition_pion_formfactor}
F_\pi( t ) = \frac{2}{3} F_\pi^u( t ) - \frac{1}{3} F_\pi^d( t ). 
\end{equation}
As a consequence of \refeq{eq:G-parity-GPD}, we derive:
\begin{equation}
F_\pi^u( t ) = \int_{-1}^{+1} \mathrm{d}x \, H^u( x, \xi, t ) = - \int_{-1}^{+1} \mathrm{d}x \, H^d( x, \xi, t ) = - F_\pi^d( t ),
\end{equation}
which means in particular that
$F_\pi( t ) = F_\pi^u( t ) = - F_\pi^d( t )$. 
This last result and \refeq{eq:H_isoscalar_both} imply the following sum rule for $H^{I=1}$:
\begin{equation}
\int_{-1}^{+1} \mathrm{d}x \, H^{I=1}(x,\xi,t) = 2 F_\pi^u( t ) = 2 F_\pi( t ),
\label{eq:sum-rule}
\end{equation}
while it imposes the vanishing of the corresponding sum rule in the isoscalar channel.

On the other hand, GPDs are even functions of $\xi$ from time reversal invariance:
\begin{equation}
  \label{eq:TimeSymmetry}
  H^{a}(x,-\xi,t)  = H^{a}(x,\xi,t),  
\end{equation}
for $a=q$ $(q = u, d)$  or $a=I$ $(I = 0, 1)$.
The physical region for $\xi$ is $[-1, +1]$ but $\xi \geq 0$ for all known processes where GPDs can be measured. In the following we will thus consider that $\xi$ is positive without loss of generality and 
write:
\begin{equation}
\cond{x^m}^a = \int_{-1}^{+1} \mathrm{d}x \, x^m H^a(x,\xi,t) \ ,
\label{eq:def_Mellin_moment}
\end{equation}
the $(m+1)^{\textrm{th}}$ Mellin moment of the GPD $H^a$. From the definition \refeq{eq-def-GPD-H-spinless-target} for $q$ = $u$, $d$, we see that:
\begin{eqnarray}
\cond{x^m}^q
& = & 
\frac{1}{2(P\cdot n)^{m+1}}\elt{\pi, P + \Dd}{\bar{q}(0)\gamma^+(i\overleftrightarrow{D}^+)^m q(0)}{\pi, P - \Dd},   \label{eq:MellinMoments}
\end{eqnarray}
where $\overleftrightarrow{D}= \frac{1}{2}\left(\overrightarrow{D}-\overleftarrow{D}\right)$ and $D$ stands for the covariant derivative applied on the left or the right hand sides. 
It is easy to see that $H$ is uniquely defined by its Mellin moments. 
\refeq{eq:MellinMoments} will be our starting point in \refsec{sec:GPD_modeling}.

Remembering \refeq{eq:H_isoscalar_both} and \refeq{eq:TimeSymmetry} the polynomiality property reads:
\begin{eqnarray}
 \int_{-1}^{1} \mathrm{d}x\, x^m H^{I=0}(x,\xi,t)
& = & 
0 \quad (m \textrm{ even}),  \label{eq:polynomiality_isoscalar_even_power} \\
\int_{-1}^{1} \mathrm{d}x\, x^m H^{I=0}(x,\xi,t) 
& = & 
\sum_{i=0 \atop {\rm even}}^{m}(2\xi )^i C_{m i}^{I=0}(t) + (2\xi )^{m+1}C_{m \, m+1}^{I=0}(t) \quad (m \textrm{ odd}),   \label{eq:polynomiality_isoscalar_odd_power} \\
\int_{-1}^{1} \mathrm{d}x\, x^m H^{I=1}(x,\xi,t) 
& = & 
\sum_{i=0 \atop {\rm even}}^{m}(2\xi )^i C_{m i}^{I=1}(t) \quad (m \textrm{ even}),   \label{eq:polynomiality_isovector_even_power} \\
 \int_{-1}^{1} \mathrm{d}x\, x^m H^{I=1}(x,\xi,t) 
& = & 
0 \quad (m \textrm{ odd}).   \label{eq:polynomiality_isovector_odd_power}
\end{eqnarray}
The coefficients $C_{m i}^I(t)$ are sometimes called generalized form factors. For a given integer $m$, \refeqs{eq:polynomiality_isoscalar_even_power}{eq:polynomiality_isovector_odd_power} establish that the highest powers of $\xi$ appear only in the isoscalar GPD. 


\subsection{Double Distributions and support properties}
\label{sec:DoubleDistributions}
 
Using notations similar to those of \refeq{eq-def-GPD-H-spinless-target}, the DDs $F^q$ and $G^q$ associated to the quark flavor $q$ are defined by the following matrix element:
\begin{eqnarray}
\bra{P+\frac{\Delta}{2}} \bar{q}\left( -\frac{z}{2} \right) \gamma^\mu q \left( \frac{z}{2} \right) \ket{P-\frac{\Delta}{2}}_{z^2=0} 
& = & 2P^{\mu}\int_{\Omega} \mathrm{d}\beta\mathrm{d}\alpha \, e^{- i \beta P \cdot z + i \alpha \frac{\Delta \cdot z}{2}} F^q( \beta, \alpha, t ) \nonumber \\
& & \, - \Delta^{\mu}\int_{\Omega} \mathrm{d}\beta\mathrm{d}\alpha \, e^{- i \beta P \cdot z + i \alpha \frac{\Delta \cdot z}{2}} G^q( \beta, \alpha, t ) \nonumber \\
& & +\text{ higher twist terms}.
\label{eq:def-DD-F-G}
\end{eqnarray}
Unless explicitly needed, the $t$-dependence will not be mentioned. $F^q$ and $G^q$ vanish outside the rhombus $\Omega$ defined by: 
\begin{equation}
\label{eq:def-rhombus-DD}
|\alpha| + |\beta| \leq 1. 
\end{equation}
The invariance of QCD under time reversal implies that $F^q$ is $\alpha$-even and $G^q$ is $\alpha$-odd.

Using the definition of the skewness $\xi$, the projection of \refeq{eq:def-DD-F-G} onto the light-cone vector $n$ writes:
\begin{eqnarray}
\bra{P+\frac{\Delta}{2}} \bar{q}\left( -\frac{z}{2} \right) \gamma^+ q \left( \frac{z}{2} \right) \ket{P-\frac{\Delta}{2}}_{z^+=0 \atop z_\perp=0} 
& = & 
2 P^+ \int_{\Omega} \mathrm{d}\beta\mathrm{d}\alpha \, e^{- i P^+ z^- ( \beta + \alpha \xi )} \big( F^q( \beta, \alpha )  + \xi G^q( \beta, \alpha ) \big). \nonumber \\
\label{eq:lightcone-projection-DD-F-G}
\end{eqnarray}
Integrating over $z^-$ the \lhs of \refeq{eq:lightcone-projection-DD-F-G} multiplied by the phase term $e^{i x P^+ z^-}$ yields the well-known relation between GPDs and DDs:
\begin{equation}
H^q( x, \xi, t ) = \int_\Omega \mathrm{d}\beta\mathrm{d}\alpha \, \delta( x - \beta - \alpha \xi ) \big( F^q( \beta, \alpha, t ) + \xi G^q( \beta, \alpha, t ) \big).
\label{eq:relation-DD-GPD}
\end{equation}
The support property $x \in [-1, +1]$ of the GPD $H^q$ can be readily obtained from this equation. 

We now introduce the twist-2 quark operator:
\begin{equation}
\label{eq:def_quark_twist_two_operator}
\mathcal{O}^{\mu\mu_1\ldots\mu_m}_q =\TwistTwoQuarkOp{\mu}{\mu_1}{\mu_m},
\end{equation}
where the notation $\{\ldots\}$ indicates complete symmetrization and trace subtraction of the enclosed indices. The expansion of the \lhs of \refeq{eq:def-DD-F-G} writes:
\begin{eqnarray}
\bra{P+\frac{\Delta}{2}} \bar{q}\left( -\frac{z}{2} \right) \gamma^\mu q \left( \frac{z}{2} \right) \ket{P-\frac{\Delta}{2}}
& = & 
\sum_{m=0}^\infty \frac{( -i )^m}{m!} z_{\mu_1} \ldots z_{\mu_m} \bra{P+\frac{\Delta}{2}} \mathcal{O}^{\mu\mu_1\ldots\mu_m}_q( 0 ) \ket{P-\frac{\Delta}{2}} \nonumber \\
& & +\text{ higher twist terms},
\label{eq:taylor-expansion-quark-bilocal-operator}
\end{eqnarray}
while the expansion of the \rhs reads:
\begin{eqnarray}
& & 
2P^{\mu}\int_{\Omega} \mathrm{d}\beta\mathrm{d}\alpha \, e^{- i \beta P \cdot z + i \alpha \frac{\Delta \cdot z}{2}} F^q( \beta, \alpha ) - \Delta^{\mu}\int_{\Omega} \mathrm{d}\beta\mathrm{d}\alpha \, e^{- i \beta P \cdot z + i \alpha \frac{\Delta \cdot z}{2}} G^q( \beta, \alpha ) \nonumber \\
& = & 
\sum_{m=0}^{\infty} \int_{\Omega} \mathrm{d}\beta\mathrm{d}\alpha \, \big( 2 F^q( \beta, \alpha ) P^\mu - \Delta^\mu G^q( \beta, \alpha ) \big) \frac{( -i )^m}{m!} \left( \beta P \cdot z - \alpha \frac{\Delta \cdot z}{2} \right)^m \nonumber \\
& & 
+\text{ higher twist terms}, \nonumber \\
& = & 
\sum_{m=0}^{\infty} \int_{\Omega} \mathrm{d}\beta\mathrm{d}\alpha \, \big( 2 F^q( \beta, \alpha ) P^\mu - \Delta^\mu G^q( \beta, \alpha ) \big) \frac{( -i )^m}{m!} \sum_{k=0}^m \binom{m}{k} \beta^{m-k} (P \cdot z)^{m-k} \alpha^k \left( - \frac{(\Delta \cdot z)}{2} \right)^k \nonumber \\
& & 
+\text{ higher twist terms}.
\label{eq:expansion-DD-phase-term}
\end{eqnarray}
Consider the moments $F^q_{mk}( t )$ and $G^q_{mk}( t )$:
\begin{eqnarray}
F^q_{mk} 
& = & \int_{\Omega} \mathrm{d}\beta\mathrm{d}\alpha \, \alpha^k \beta^{m-k} F^q( \beta, \alpha ), \label{eq:def-moment-DD-F} \\
G^q_{mk} 
& = & \int_{\Omega} \mathrm{d}\beta\mathrm{d}\alpha \, \alpha^k \beta^{m-k} G^q( \beta, \alpha ). \label{eq:def-moment-DD-G}
\end{eqnarray}
The identification of leading-twist terms in both \refeq{eq:taylor-expansion-quark-bilocal-operator} and (\ref{eq:expansion-DD-phase-term}) yields:
\begin{eqnarray}
\bra{P+\frac{\Delta}{2}} \mathcal{O}^{\mu\mu_1\ldots\mu_m}_q( 0 ) \ket{P-\frac{\Delta}{2}} & = & \sum_{k=0}^m \binom{m}{k} \big[ F^q_{mk}( t )  2 P^{\{\mu} - G^q_{mk}( t )  \Delta^{\{\mu} \big] P^{\mu_1} \ldots P^{\mu_{m-k}} \nonumber \\
& & \left( -\frac{\Delta}{2}  \right)^{\mu_{m-k+1}} \ldots \left( -\frac{\Delta}{2} \right)^{\mu_m\}}.
\label{eq:twist-two-quark-operator-def-DD}
\end{eqnarray}
\refeq{eq:twist-two-quark-operator-def-DD} will be the key element for the determination of the DDs from the matrix elements between pion states of the twist-2 quark operators.


\section{GPD modeling in the \ds~--~\bs approach}
\label{sec:GPD_modeling}

Modeling GPDs remains today a hard task and must be done carefully. Indeed \refeq{eq-def-GPD-H-spinless-target} shows that GPDs are nonlocal nonperturbative objects defined on the light cone of Minkowskian spacetime. GPDs models should pass stringent tests, like the fulfillment of the polynomiality property, which is at the same time hard to implement and  not constraining enough to pin down a first principle parameterization.

As emphasized above, a GPD is uniquely defined by its Mellin moments. The relation (\ref{eq:MellinMoments}) expresses the Mellin moments of a GPD in terms of matrix elements of local operators; this allows the computation of Mellin moments in Euclidean spacetime (as it is the case in the \ds or lattice QCD approaches) and the translation of the result back to Minkowskian spacetime. We have also seen that the interplay of Lorentz covariance and discrete symmetry is most visible in the isoscalar and isovector GPDs through \refeqs{eq:polynomiality_isoscalar_even_power}{eq:polynomiality_isovector_odd_power}. 

However the reconstruction of the GPD from the knowledge of its Mellin moments is a nontrivial task that will be further discussed in a future publication. This problem can be solved by computing DDs since the relation (\ref{eq:relation-DD-GPD}) allows a direct reconstruction of the GPD. The matrix element involved in the computation (\ref{eq:MellinMoments}) of the Mellin moments of the GPD is parameterized in terms of DDs as written down in \refeq{eq:twist-two-quark-operator-def-DD}.

\begin{figure}[htb]
	\centering
	\begin{tabular}{cc}
  	\includegraphics[width=0.30\textwidth]{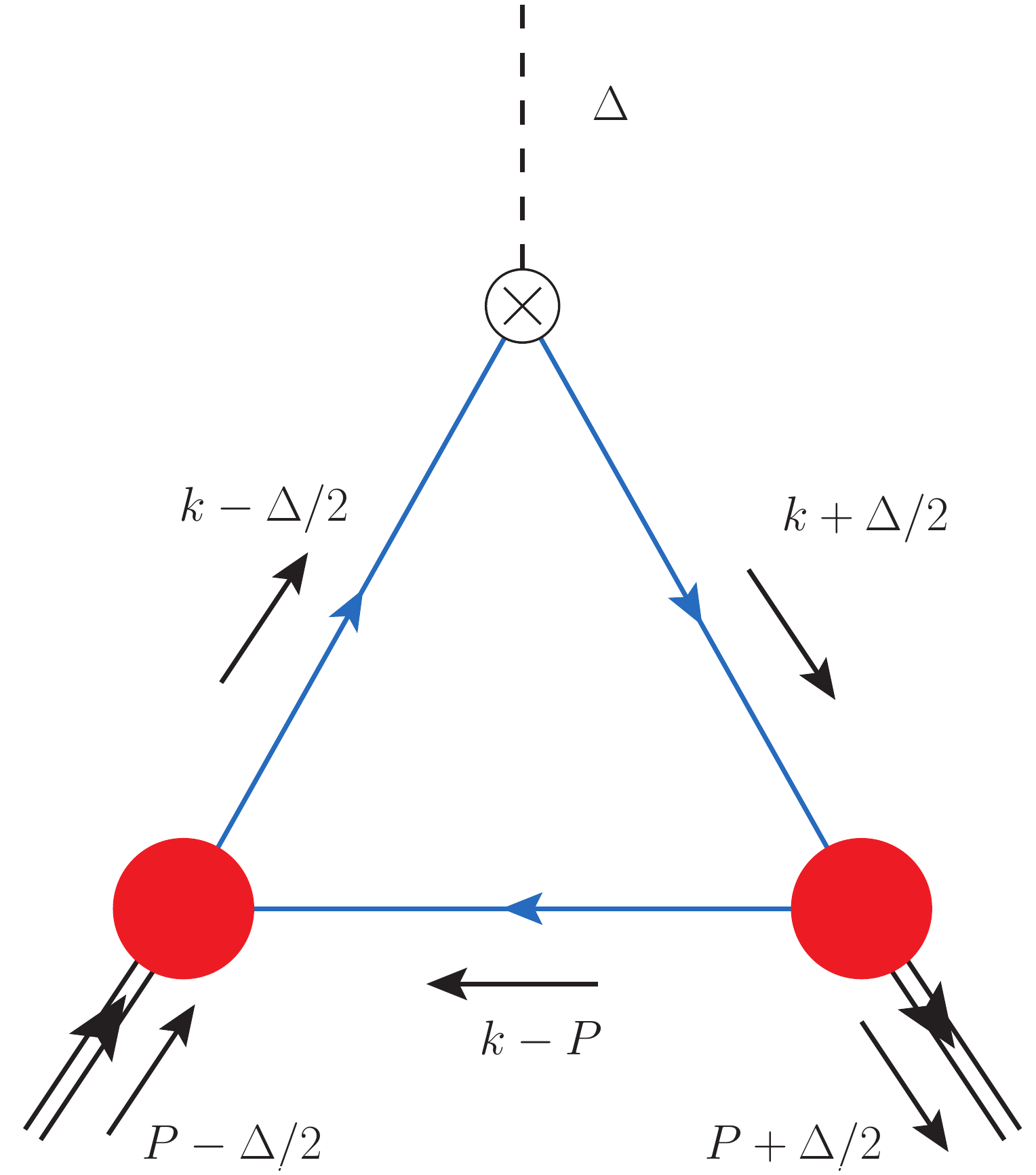} & \rule[0cm]{1cm}{0cm}
  	\includegraphics[width=0.32\textwidth]{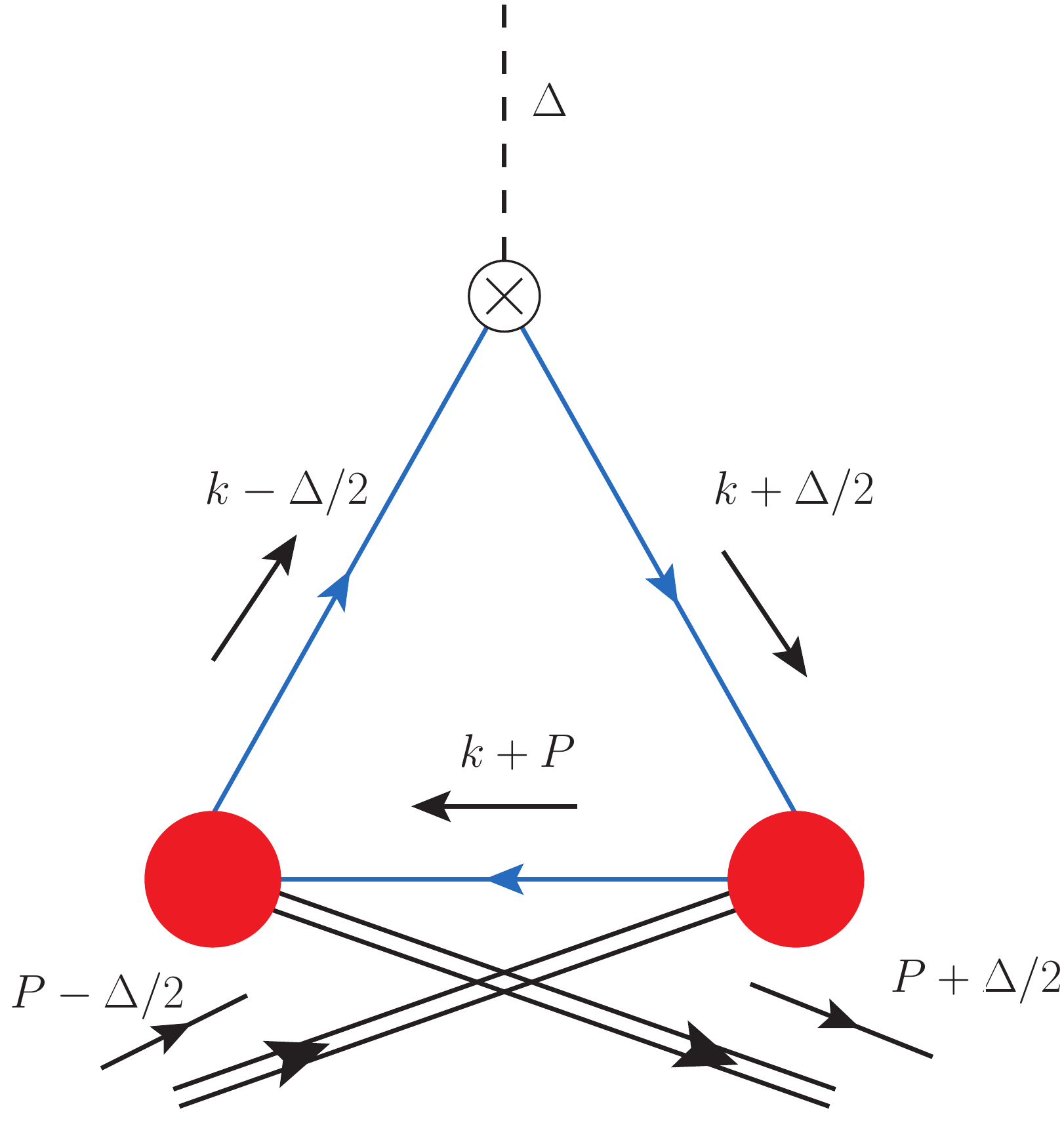}
  	\end{tabular}
  	\caption{\small Triangle diagram computation of the Mellin moments of the pion GPD $H$. In blue, the propagators taken into account in the computations. The crosses represent the insertions of the tower of twist-2 quark operators $\mathcal{O}^{\mu\mu_1\ldots\mu_j}_q$ (\ref{eq:def_quark_twist_two_operator}) with incoming 4-momentum $\Delta$.}
 	\label{fig:TriangleDiagrams}
\end{figure}

Our modeling strategy thus consists in the identification of the DDs $F^q$ and $G^q$ from the computation of the matrix element between pion states of momenta $P \pm \Delta/2$ of the twist-2 quark operator $\mathcal{O}^{\mu\mu_1\ldots\mu_m}_q$, as can be diagramatically seen in Fig.~\ref{fig:TriangleDiagrams}. From now on, all definitions and computations will be expressed in Euclidean spacetime, and we will come back to Minkowskian spacetime in \refsec{sec:results} to compare our model to existing data.

\begin{figure}[htb]
  \centering
  \includegraphics[width = 0.50\textwidth]{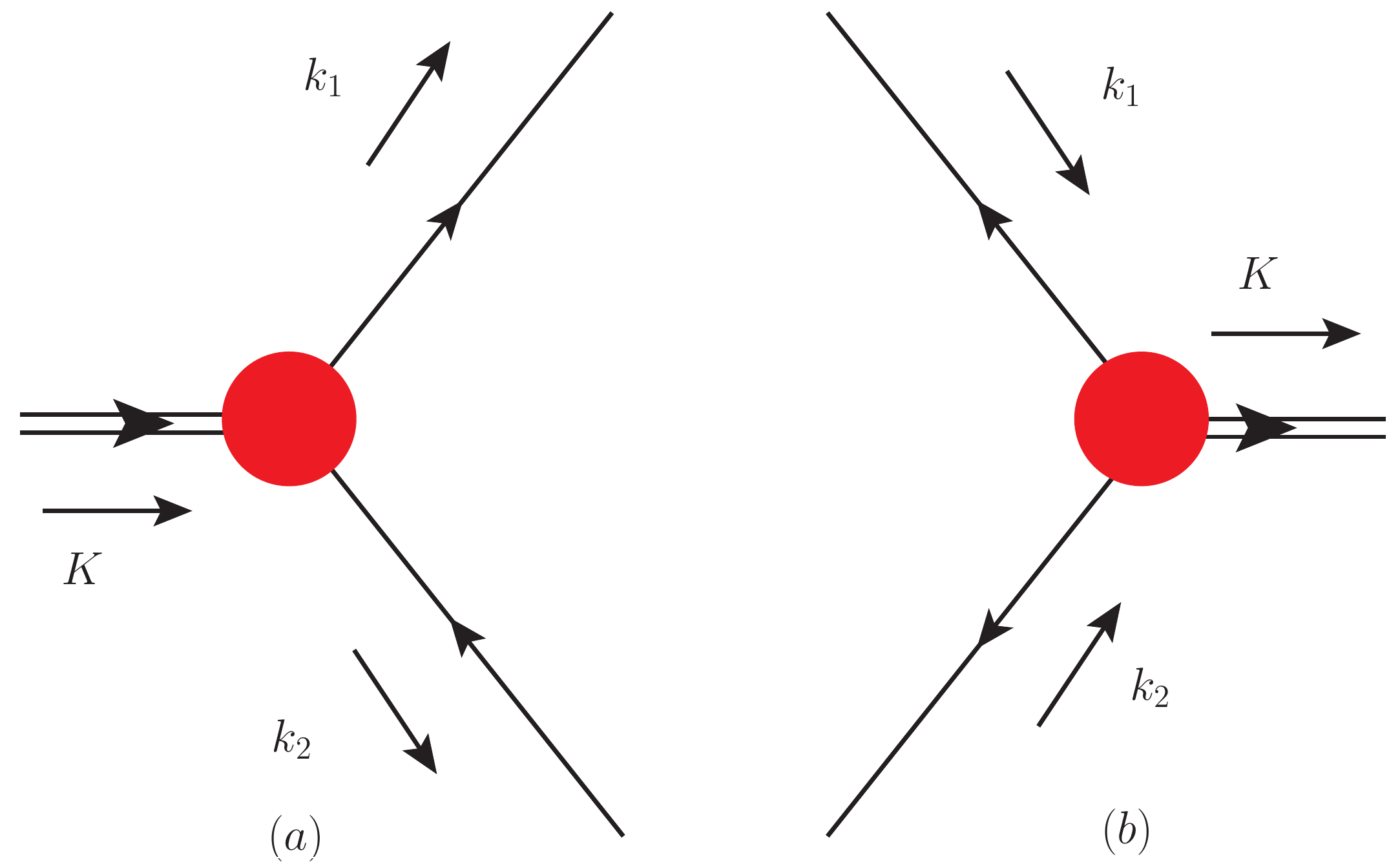}
  \caption{\small Momenta configurations for the effective pion-quark vertex (a) and its conjugate (b), defined from the \bs amplitude.}  
  \label{fig:BSA}
\end{figure}

Among the many attempts to apply the \ds formalism to describe the structure of hadrons (see, for instance \refcite{Maris:2003vk,Bashir:2012fs} and references therein), 
we take \refcite{Chang:2013pq} as a enlightening starting point. There, the authors developed a systematic procedure to compute the pion DA from the \bs amplitude. Their model relies on the computation of the DA's Mellin moments supplemented by an appropriate reconstruction method. This strategy is adapted in the present study to the computation of the pion GPD $H$.
In spite of the phenomenological relevance of gluon GPDs \cite{Moutarde:2013qs} or the general interest in gluon contributions by themselves \cite{Bashir:2013zha}, we model only quark GPDs. We further neglect gluon contributions that are not included in the gap equations, the \dse or the evolution equations for QCD. This may be harmless for the phenomenology of pion GPDs since most of the relevant data lie in the valence region \cite{Conway:1989fs,Bordalo:1987cr,Betev:1985pf}. Thus, we write the $(m+1)^{\textrm{th}}$ Mellin moment of $H$ 
as\footnote{As discussed in \refsec{sec:GPD_Definition}, it is enough to compute the GPDs corresponding to the $\pi^+$ state, so the \bs vertices are accompanied by the matrices $\tau_\pm$. From isospin symmetry we restrict ourselves to the computation of $H^u$ and $H^d$, or simply of $H^u(x,\xi,t)$ and $H^u(-x,\xi,t)$ ($0\leq x \leq 1$) thanks to \refeq{eq:G-parity-GPD} (one function corresponding to a diagram, and the other to its crossed version).}:
\begin{eqnarray}
\label{eq:TriangleDiagrams} 
2 ( \Pn )^{m+1} \cond{x^{m}}^u & = &  \mathrm{tr}_{CFD} \int \frac{\mathrm{d}^4k}{(2\pi)^4}(k\cdot n)^m \, \tau_+ i\Gamma_\pi\left(\eta(k-P)+(1-\eta)\left(k-\Dd\right),  P-\Dd \right)\nonumber \\
& &  S( k - \Dd ) \ i\gamma^+~ S( k+\Dd ) \nonumber \\
& & \tau_- i\bar{\Gamma}_\pi\left((1-\eta)\left(k+\Dd\right)+\eta(k-P),P+\Dd \right) S( k - P ),
\end{eqnarray}
where $\mathrm{tr}_{CFD}$ indicates that our expression is traced on color, isospin, and Dirac indices and the $\tau$'s are isospin matrices  (see Figs.~\ref{fig:TriangleDiagrams} and \ref{fig:BSA}). In \refeq{eq:TriangleDiagrams}, $S$ stands for the fully dressed renormalized quark propagator, and $\Gamma_\pi$ is the effective pion-quark vertex
which can be written in terms of the \bs amplitude, $\chi$, as
\begin{equation}
  \label{eq:PionQuarkVertex}
  \G_{\pi}(k,K) = S^{-1}(k_2)\ \chi(k,K) \ S^{-1}(k_1) ,
\end{equation}
where $K = k_1+k_2$ and $k = (1-\eta) k_1 - \eta k_2$. $k_1$ and $k_2$ stand for the quark momenta leaving the vertex. The parameter $\eta \in [0,1]$ describes 
the arbitrariness in defining the position of the center of mass of the $q\bar{q}$ pair. Owing to Poincar\'e covariance, the final result does not depend on it. 
The conjugate  pion-quark vertex, $\bar{\Gamma}_\pi$, writes:
\begin{equation}
  \label{eq:PionBarVertex}
  \bar{\Gamma}_{\pi}(k,K) = C^{\dagger} \Gamma^T_{\bar{\pi}}(-k,-K) C,
\end{equation}
where $C=\g^0 \g^2$ is the charge conjugation matrix and the superscript $T$ denotes transposition. 
Momentum flows and conventions are pictured in \reffig{fig:BSA}.

As explained in \refcite{Chang:2013pq}, both the \bs amplitude and the full quark propagator can be  computed nonpertubatively by solving the corresponding gap and \bs equations. This requires an appropriate truncation scheme and allows to obtain enough Mellin moments to reconstruct the pion DA. However, in this paper, as a valuable first step, we will apply the following simple algebraic model for both the euclidean quark propagator and the \bs amplitude:
\begin{eqnarray}
S( p ) 
& = & \big[ - i \gamma \cdot p + M \big] \Delta_M( p ^2 ), \label{eq:toy-model-quark-propagator} \\
\Delta_M( s )
& = & \frac{1}{s + M^2}, \label{eq:toy-model-propagator-mass-term} \\
\Gamma_\pi( k, p )
& = & i \gamma_5 \frac{M}{f_\pi}M^{2\nu} \int_{-1}^{+1} \mathrm{d}z \, \rho_\nu( z ) \ 
\left[\Delta_M( k_{+z}^2 )\right]^\nu; \label{eq:toy-model-bs-vertex} \\
\rho_\nu( z ) & = & R_\nu ( 1 - z^2 )^\nu, \label{eq:toy-model-vertex-smearing-function} 
\end{eqnarray}
where $R_\nu$ normalizes to 1 the integral of $\rho_\nu$ over $z \in [-1, +1]$ and $k_{+z}=k-\left(\frac{1-z}{2}-\eta \right)p$. $M$ is the only dimensionful parameter of the model and appears as an effective quark mass. This model has the merit of exhibiting most of the features of a realistic computation involving the numerical solutions of the \bs and \ds equations, but being easier to elucidate. It will allow us to control our work and assumptions step by step. To give the reader a quantitative understanding of the variable $\nu$, let us remind that in \refcite{Chang:2013pq}, $\nu = 0$ corresponds to a flat DA while $\nu = 1$ describes an asymptotic DA.

Concerning the normalization for the \bs amplitude, 
Mandelstam \cite{Mandelstam:1955sd} proposed a normalization condition relying on charge conservation in the ladder approximation. It amounts to compute a form factor at vanishing momentum transfer with a triangle diagram approximation (as in \refeq{eq:TriangleDiagrams} for $m$ = 0). 

Let us note $\Gamma^{\textrm{e.m.}}_\mu$ the quark-quark-photon vertex. Using the Ward~-~Takahashi identity:
\begin{equation}
i \Delta^\mu \Gamma^{\textrm{e.m.}}_\mu(k+\frac{\Delta}2,k-\frac{\Delta}2) \ = \ S^{-1}(k+\frac{\Delta}2) - S^{-1}(k-\frac{\Delta}2), 
\label{eq:WT-identity}
\end{equation}
Mandelstam's condition was shown \cite{Lurie:1965,Nishijima:1967} to be equivalent to the canonical normalization of the \bs amplitude, in which the 2-quark amplitude has unit residue at the bound state pole. 

In our algebraic model, the parameterization of the quark propagator \refeqs{eq:toy-model-quark-propagator}{eq:toy-model-propagator-mass-term} automatically fulfills the condition (\ref{eq:WT-identity}) with the bare vertex $\gamma_\mu$. Thus our Ansatz \refeqs{eq:toy-model-bs-vertex}{eq:toy-model-vertex-smearing-function} for the \bs amplitude will be canonically normalized if we impose $\cond{x^{m}}^{I=1} = 2$ when $t = 0$.

We will thus inject \refeqs{eq:toy-model-quark-propagator}{eq:toy-model-vertex-smearing-function} into \refeq{eq:TriangleDiagrams}, for $\eta=0$, in order to generate the Mellin's moments for the pion GPD.


\section{Results: theoretical constraints and phenomenology}
\label{sec:results}

Hereafter, our purpose is the evaluation of \refeq{eq:TriangleDiagrams} with \refeqs{eq:toy-model-quark-propagator}{eq:toy-model-vertex-smearing-function} for 
the \bs amplitude and quark propagators, in the chiral limit $m_\pi \rightarrow 0$ ($m_\pi^2 \ll M^2$, $M$ being the only dimensionful scale of our algebraic model). First we check the expected constraints on GPDs from general theoretical arguments. Then we proceed to the reconstruction of the pion PDF from its Mellin moments and to the computation of the pion form factor. Finally we compare with experimental results.



\subsection{Support, polynomiality and discrete symmetries}

First we outline the evaluation of the leftmost ("direct") Feynman graph in \reffig{fig:TriangleDiagrams}, giving access to the valence GPD $H^u$; the computation of the rightmost ("crossed") Feynman graph, related to $H^d$, is the same \textit{mutatis mutandis}. Starting from \refeq{eq:TriangleDiagrams}, one should first replace the full quark propagator and effective pion-quark vertex 
by \refeqs{eq:toy-model-quark-propagator}{eq:toy-model-vertex-smearing-function} and use \refeq{eq:PionBarVertex} for the conjugate pion-quark vertex.

The traces on flavor and color indices merely produce an overall factor. To evaluate the loop integral over $k$, we extend the approach of  \refcite{Tiburzi:2002tq} and get rid of the terms linear in $k$ by introducing explicitly the denominators of the propagators of the triangle diagrams. Namely the trace on Dirac indices give the following structure to the numerator of \refeq{eq:TriangleDiagrams}: 
\begin{eqnarray}
  \label{eq:dirac_traces}
   & &\textrm{Tr}\left( i \gamma_5 \left[-i \left( k-\Dd \right) \cdot \gamma + M \right] \gamma^{\mu} \left[ -i \left( k+\Dd \right) \cdot \gamma + M \right] i \gamma_5 \big[ -i ( k-P ) \cdot \gamma + M \big] \right) \nonumber \\
   & = & 4i\left[ k^\mu \left( k^2 - 2k \cdot P + M^2 + \left( \Dd \right)^2 \right) + P^\mu \left( k^2 - \left( \Dd \right)^2 + M^2 \right) - \Dd^\mu 2k \cdot \Dd \right]. \nonumber \\
   & = & 4i \left[ k^\mu \left( C + \left( \Dd \right)^2 - P^2 \right) + \frac{P^\mu}{2} \left( A + B - 4 \left( \Dd \right)^2 \right) - \frac{\Delta^\mu}{2} \left( \frac{B-A}{2} \right) \right],
\end{eqnarray}
with $A = \left( k - \Dd \right)^2 + M^2$, $B = \left( k + \Dd \right )^2 + M^2$, and $C = ( k - P )^2 + M^2$. Introducing Feynman parameters $x, y, u, v, w \in [0, 1]$, one can integrate out the $k$ dependence. In order to shorten the equations, we set: 
\begin{eqnarray}
  \label{eq:shortcut_def}
  f(x,y,v,w,z,z') & = & \frac{1}{2} \left( -\frac{1+z'}{2}y + \frac{1+z}{2}x + v-w \right), \\
  g(x,y,u,z,z') & = & \left(\frac{1-z'}{2}\right ) y + x \frac{1-z}{2} + u,   \\
  M'(t,P^2,x,y,u,v,w,z,z')^2 & = & M^2 +\frac{t}{4}\left( -4f^2 + y\left( \frac{1+z'}{2}\right)^2 + x \left( \frac{1+z}{2} \right )^2 +v + w \right ) \nonumber \\  
& &  +P^2 \left( -g^2 + \left(\frac{1-z'}{2}\right)^2 y + \left(\frac{1-z}{2}\right)^2 x + u \right ).
\end{eqnarray}
Our result for any Mellin moment involves the following matrix element:
\begin{eqnarray}
&&\bra{P+\frac{\Delta}{2}} \mathcal{O}^{\mu\mu_1\ldots\mu_m}_u( 0 ) \ket{P-\frac{\Delta}{2}}_{\textrm{direct}}
= \nonumber \\
&&\lambda \int_0^1 \mathrm{d}x \, \mathrm{d}y \, \mathrm{d}u \, \mathrm{d}v \, \mathrm{d}w \,  
\int_{-1}^{+1} \mathrm{d}z \, \mathrm{d}z' \, \delta( x + y + u + v + w - 1 ) ( x y )^{\nu - 1}\rho( z ) \rho( z' ) \nonumber \\
&& \frac{M^{4\nu}}{2} \left[ \frac{\Gamma( 2 \nu + 1 )}{\Gamma( \nu )^2} \left( \big( f \Delta^{\{\mu} + g P^{\{\mu} \left( \left( \Dd \right)^2 - P^2 \right) - 2 P^{\{\mu} \left( \Dd \right)^2 \right) \frac{1}{(M')^{2 \nu + 1 }} \right. \nonumber \\
&&+ \frac{\Gamma( 2 \nu )}{\Gamma( \nu )^2} \frac{1}{2} \left( P^{\{\mu} + \Dd^{\{\mu} \right) \delta( v ) \frac{1}{(M')^{2\nu}}+ \frac{\Gamma( 2 \nu )}{\Gamma( \nu )^2} \frac{1}{2} \left( P^{\{\mu} - \Dd^{\{\mu} \right) \delta( w ) \frac{1}{(M')^{2\nu}} \nonumber \\
&&\left. 
+ \frac{\Gamma( 2 \nu )}{\Gamma( \nu )^2} \left( f \Delta^{\{\mu} + g P^{\{\mu} \right) \delta( u ) \frac{1}{(M')^{2\nu}} \right] ( f \Delta + g P )^{\mu_1} \ldots ( f \Delta + g P )^{\mu_m\}},
\label{eq:mellin-moment-direct-diagram-dd}
\end{eqnarray}
where $\lambda$ is a normalization constant.
The computation of the crossed diagram can be performed along the same lines. The final expression for the Mellin moments of the isovector and isoscalar GPDs is given in \refapp{app:A}.

The DDs $F^u$ and $G^u$ are then obtained by inspection of \refeq{eq:mellin-moment-direct-diagram-dd} using the following change of variables:
\begin{equation}
\int_0^1 \mathrm{d}x \, \mathrm{d}y \, \mathrm{d}u \, \mathrm{d}v \, \mathrm{d}w \, \int_{-1}^{+1} \mathrm{d}z \, \mathrm{d}z' \, \delta( x + y + u + v + w - 1 ) \phi( x, y, u, v, w, z, z' ) = \int_{\Omega} \mathrm{d}\beta \, \mathrm{d}\alpha \Phi( \beta, \alpha ),
\label{eq:change-of-variable-rhombus}
\end{equation}
with:
\begin{eqnarray}
\Phi( \beta, \alpha ) & = & \frac{1}{16} \int_{\beta+\alpha}^{+1} \mathrm{d}B \int_{\beta-\alpha}^{+1} \mathrm{d}B' \,  \int_{-1}^{\beta+\alpha} \mathrm{d}A \, \int_{-1}^{\beta-\alpha} \mathrm{d}A'  \, \theta( A + A' ) \frac{1}{( B - A ) ( B' - A' )} \nonumber \\
& & \phi\left( \frac{-A + B}{2}, \frac{B' - A'}{2}, \frac{A + A'}{2}, \frac{1 - B}{2}, \frac{1 - B'}{2},\frac{- ( A + B ) + 2 ( \beta + \alpha )}{A - B},  \frac{- ( A' + B' ) + 2 ( \beta - \alpha )}{A' - B'} \right), \nonumber \\
\end{eqnarray}
and:
\begin{eqnarray}
\alpha & = & - f( x, y, v, w, z, z' ), \\
\beta & = & g( x, y, u, z, z' ). 
\end{eqnarray}
To demonstrate that the variables $\alpha$ and $\beta$ indeed live in the rhombus $\Omega$, we introduce a system of barycentric coordinates $(x_i)_{1 \leq i \leq 4}$ in $[0, 1]$ such that:
\begin{eqnarray}
x & = & x_4, \label{eq:substx} \\
y & = & x_3 ( 1 - x_4 ), \label{eq:substy} \\
u & = & x_2 ( 1 - x_3 ) ( 1 - x_4 ), \label{eq:substu} \\
v & = & x_1 ( 1 - x_2 ) ( 1 - x_3 ) ( 1 - x_4 ), \label{eq:substv} \\
w & = & ( 1 - x_1 )  ( 1 - x_2 ) ( 1 - x_3 ) ( 1 - x_4 ). \label{eq:substw} 
\end{eqnarray}
Using these new variables, the expressions of $\beta - \alpha$ and $\beta+\alpha$ read:
\begin{eqnarray}
\beta + \alpha
& = & 1 - \big( x_4 ( 1 + z ) + ( 1 - x_4 ) [ 2 x_1 ( 1 - x_2 ) ( 1 - x_3 ) ] \big), \label{eq:alphap} \\
\beta - \alpha
& = & - 1 + 2 \left( x_4 + ( 1 - x_4 ) \left[ x_3 \frac{1-z'}{2} + ( 1 - x_3 ) \big( x_2 + ( 1 - x_2 ) x_1 \big) \right] \right).    \label{eq:betap} 
\end{eqnarray}
In \refeq{eq:alphap} we recognize the center of mass of the system $(1 + z )$ and $[ 2 x_1 ( 1 - x_2 ) ( 1 - x_3 ) ]$ with respective weights $x_4$ and $1 - x_4$, which means that $x_4 ( 1 + z ) + ( 1 - x_4 ) [ 2 x_1 ( 1 - x_2 ) ( 1 - x_3 )]$ can be any number between 0 and 2. Consequently $-1 \leq \beta + \alpha \leq +1$. The same barycentric interpretation applied to \refeq{eq:betap} establishes $-1 \leq \beta - \alpha \leq +1$. This set of inequalities can be summarized by $| \alpha | + | \beta | \leq 1$.

In passing, we note that $\beta \geq 0$ for the direct diagram, while $\beta \leq 0$ for the crossed diagrams. From \refeq{eq:relation-DD-GPD} we see that the valence GPD computed in the crossed diagram has a support $x \in [-\xi, +1]$, which is exactly what is expected on general grounds \cite{Diehl:2003ny,Belitsky:2005qn}. Note also that the same change of variables can be particularized to the direct (\ie without evaluating a GPD as an intermediate step) computation of a meson PDF. 

The DDs $F^u$ and $G^u$ we identify in our computation are respectively $\alpha$-even and $\alpha$-odd as requested from discrete symmetry requirements. To complete our calculation we only need to keep track of the normalization of the \bs amplitude. From \refeq{eq:twist-two-quark-operator-def-DD} (or its realization \refeq{eq:mellin-moment-direct-diagram-dd}) it is easy to compute the Mellin moments of the GPD $H^u$ and use the normalization condition (\ref{eq:sum-rule}) at $t = 0$, hence specifying the multiplicative factor $\lambda$ (see \refapp{app:A}). Within the previously developed framework, we are able to compute analytically every integral and thus to give analytic expressions of the DDs $F^u$ and $G^u$, as shown in \refapp{app:A}. 

Let us summarize briefly. Starting with a triangle diagram evaluation, five Feynman parameters $x, y, u, v, w$ (living in $[0, 1]$) and two convolution parameters $z, z'$ (living in $[-1, +1]$), we have analytically demonstrated that the support property holds. The quark longitudinal momentum fraction is smaller than 1 and the behavior of DDs at the cusps of the rhombus ensures the vanishing of the GPD at $x = \pm 1$ and the continuity of the GPD at $x = \xi$. We can also analytically prove that the polynomiality property holds up to the highest order. 

\begin{figure}
  \begin{minipage}[b]{.46\linewidth}
    \centering
    \includegraphics[width=\linewidth]{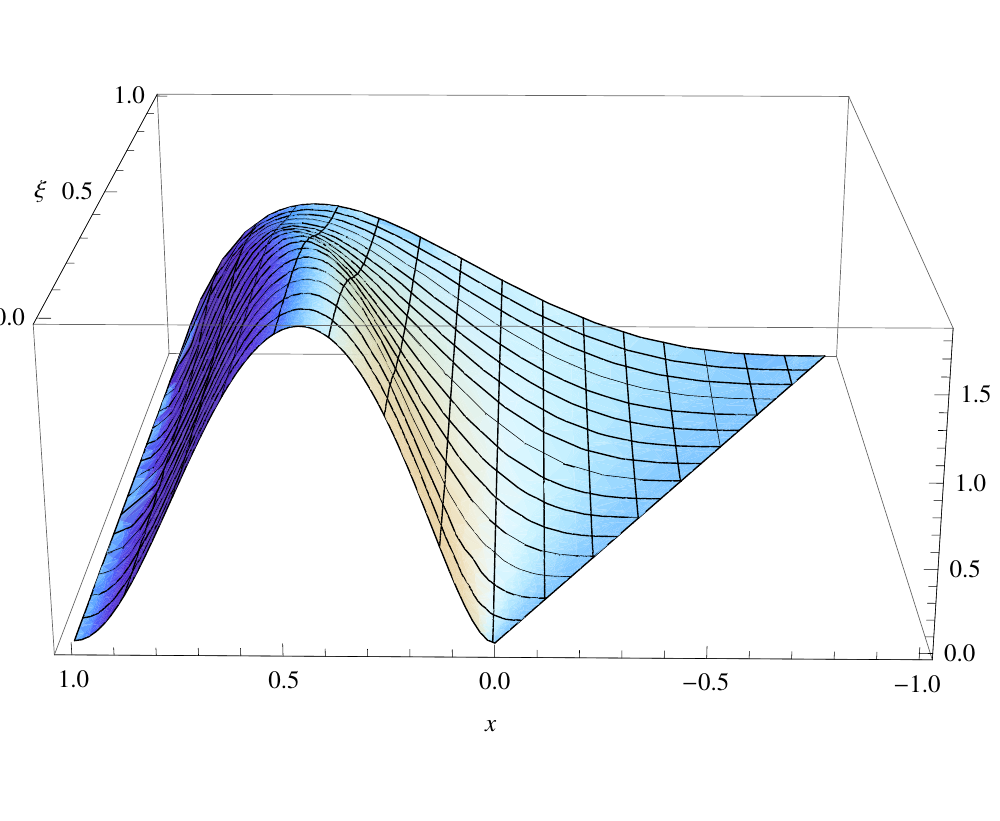}
    \caption{\small Pion valence GPD $H^u( x, \xi, t )$ as a function of $x$ and $\xi$ at vanishing $t$.}
    \label{fig:3D-pion-GPD}
  \end{minipage} \hfill
  \begin{minipage}[b]{.46\linewidth}
    \centering
    \includegraphics[width=\linewidth]{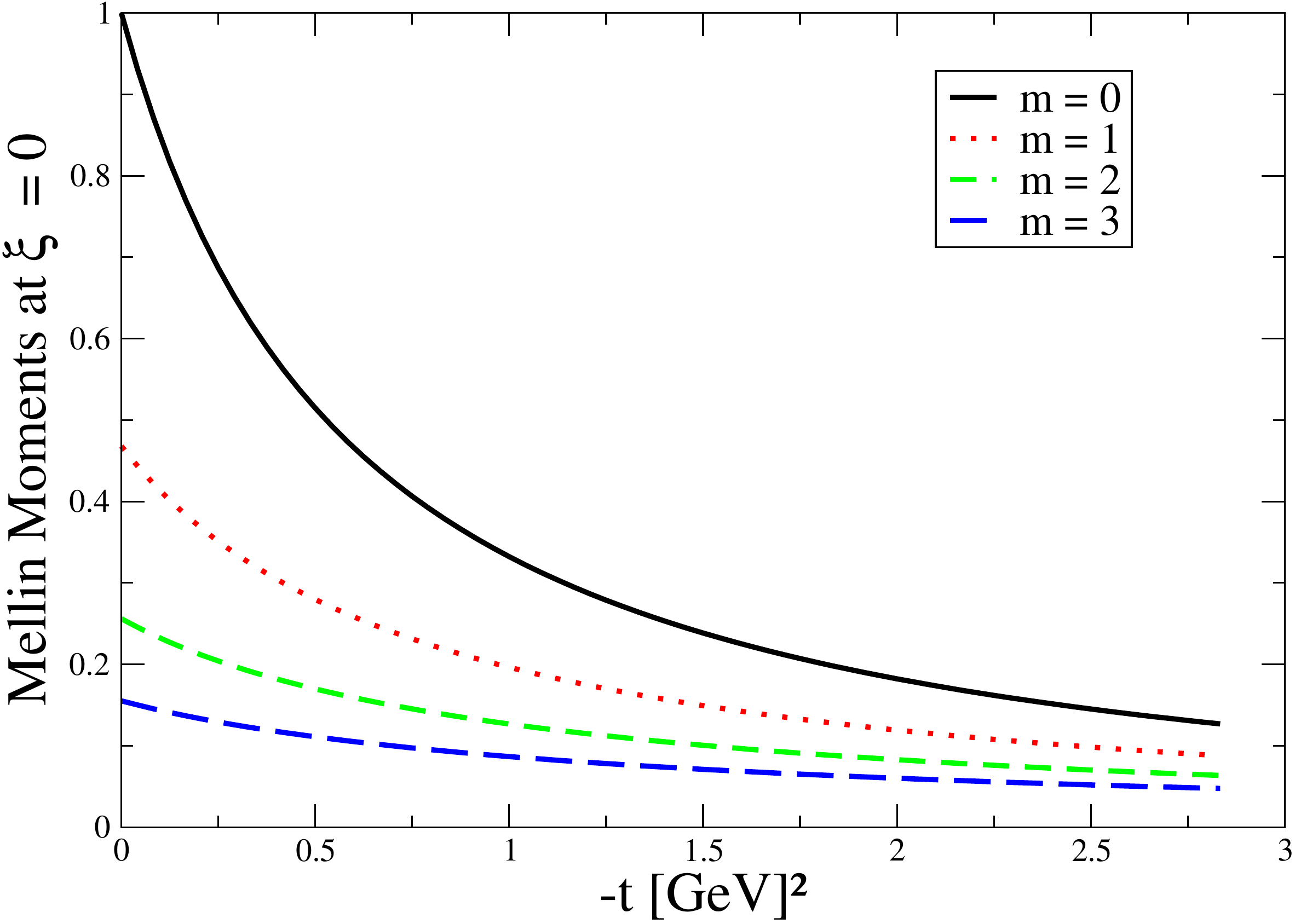}
    \caption{\small Mellin Moments of $H^u$ computed with the algebraic model at $\xi = 0$.}
    \label{fig:MellinMoments}
  \end{minipage}
\end{figure}

From now on, we will neglect the pion mass effects, \ie $P^2 \approx -\frac{t}{4}$. We will also set the single dimensionful parameter of our model, $M$, to a typical constituent quark mass: 350~\MeV. The dimensionless parameter $\nu$ is set to 1. The functional form of the GPD is shown on \reffig{fig:3D-pion-GPD}. The support property $x \in [-\xi, +1]$ is manifest.

Our full computation (see results in \refapp{app:A}) gives a direct $t$-dependence as a rational function illustrated on \reffig{fig:MellinMoments} for $\xi = 0$. On the other hand, we can take \refeqs{eq:DDF}{eq:DDG} at vanishing $t$, integrate over $\alpha$ and $\beta$ and obtain the following expression for the GPD $H^u$ in the DGLAP region:
\small
\begin{eqnarray}
  \label{eq:HDGLAP}
   H^u_{x \ge \xi}(x,\xi,0) & = &  \frac{48}{5}\left\{\frac{3 \left(-2 (x-1)^4 \left(2 x^2-5 \xi ^2+3\right) \log (1-x)\right)}{20 \left(\xi ^2-1\right)^3}\right. \nonumber \\
  & &  \frac{3 \left(+4 \xi  \left(15 x^2 (x+3)+(19 x+29) \xi ^4+5 (x (x (x+11)+21)+3) \xi ^2\right) \tanh ^{-1}\left(\frac{(x-1) \xi }{x-\xi ^2}\right)\right)}{20 \left(\xi ^2-1\right)^3} \nonumber \\
  & & +\frac{3 \left(x^3 (x (2 (x-4)x+15)-30)-15 (2 x (x+5)+5) \xi ^4\right) \log \left(x^2-\xi ^2\right)}{20 \left(\xi ^2-1\right)^3} \nonumber \\
  & & +\frac{3\left(-5 x (x (x (x+2)+36)+18) \xi ^2-15 \xi ^6\right) \log \left(x^2-\xi ^2\right)}{20 \left(\xi ^2-1\right)^3} \nonumber \\
 & & +\frac{3 \left(2 (x-1) \left((23 x+58) \xi ^4+(x (x (x+67)+112)+6) \xi ^2+x (x ((5-2 x) x+15)+3)\right)\right)}{20 \left(\xi ^2-1\right)^3}\nonumber \\
 & & +\frac{3\left(\left(15 (2 x(x+5)+5) \xi ^4+10 x (3 x (x+5)+11) \xi ^2\right) \log \left(1-\xi ^2\right)\right)}{20 \left(\xi ^2-1\right)^3} \nonumber \\
 & &+\left.\frac{3\left(2 x (5 x (x+2)-6) +15 \xi ^6-5 \xi ^2+3\right) \log \left(1-\xi ^2\right)}{20 \left(\xi ^2-1\right)^3}\right\},
\end{eqnarray}
\normalsize
and in the ERBL region:
\small
\begin{eqnarray}
  \label{eq:HERBL}
  H^u_{|x| \le \xi}(x,\xi,0) & = &  \frac{48}{5}\left\{\frac{ 6\xi (x-1)^4 \left(-\left(2 x^2-5 \xi ^2+3\right)\right) \log (1-x)}{40 \xi  \left(\xi ^2-1\right)^3}\right. \nonumber \\
 & & +\frac{6\xi \left(-4 \xi  \left(15 x^2 (x+3)+(19 x+29) \xi ^4+5 (x (x (x+11)+21)+3) \xi ^2\right) \log (2 \xi )\right)}{40 \xi  \left(\xi ^2-1\right)^3} \nonumber \\
 & & +\frac{6\xi (\xi +1)^3 \left((38 x+13) \xi ^2+6 x (5 x+6) \xi +2 x (5 x (x+2)-6)+15 \xi ^3-9 \xi +3\right) \log (\xi +1)}{40 \xi  \left(\xi ^2-1\right)^3} \nonumber \\
 & & +\frac{6\xi(x-\xi )^3 \left((7 x-58) \xi ^2+6 (x-4) x \xi +x (2 (x-4) x+15)+15 \xi ^3+75 \xi -30\right) \log (\xi -x)}{40 \xi \left(\xi ^2-1\right)^3} \nonumber \\
 & & +\frac{3(\xi -1) (x+\xi ) \left(4 x^4 \xi-2 x^3 \xi  (\xi +7)+x^2 (\xi  ((119-25 \xi ) \xi -5)+15)\right)}{40 \xi  \left(\xi ^2-1\right)^3} \nonumber \\
 & & +\left.\frac{3(\xi -1) (x+\xi)\left(x \xi  (\xi  (\xi  (71 \xi +5)+219)+9)+2 \xi  (\xi  (2 \xi  (34 \xi +5)+9)+3)\right)}{40 \xi  \left(\xi ^2-1\right)^3}\right\}.
\end{eqnarray}
\normalsize
We remind that $\xi \geq 0$. Despite apparent singularities at $\xi =$ 0 and 1 in \refeq{eq:HDGLAP} and (\ref{eq:HERBL}), it should be stressed that the GPD is actually nonsingular at these points.

\subsection{Pion Form Factor}

As our computations have been done in euclidean theory, the following comparisons require to move our kinematic variables from euclidean space to minkowskian space, \ie $t_{E}=-t_{M}$.

The sum rule (\ref{eq:sum-rule}) can now be invoked for a first phenomenological test of our simple model. Thus, one can directly compare the $t$-behavior of the pion form factor provided by our model to experimental data. 

Measurements of space-like pion electromagnetic form factors are mostly contained in two datasets. The first one was obtained by the NA7 Collaboration at CERN \cite{Amendolia:1986wj} and cover the range $0.014 < -t < 0.26~\GeV^2$ by scattering 300~\GeV\xspace pions on the electrons of a liquid hydrogen target. The measured electric charge radius of the pion is:
\begin{equation}
\label{eq:pion-electric-charge-radius-na7}
\cond{r_\pi^2}^{\textrm{exp}} = \left. -6 \frac{dF_\pi}{dt}\right|_{t=0}= 0.439 \pm 0.008~\textrm{fm}^2.
\end{equation}
The second one is a result of the $F_\pi$ Collaboration at Jefferson Lab \cite{Huber:2008id} and explores a complementary kinematic domain $0.60 < -t < 2.45~\GeV^2$ through the high-energy electroproduction of a pion on a nucleon. 

\begin{figure}[h]
  \centering
  \begin{tabular}{cc}
  \includegraphics[width= 0.48\textwidth]{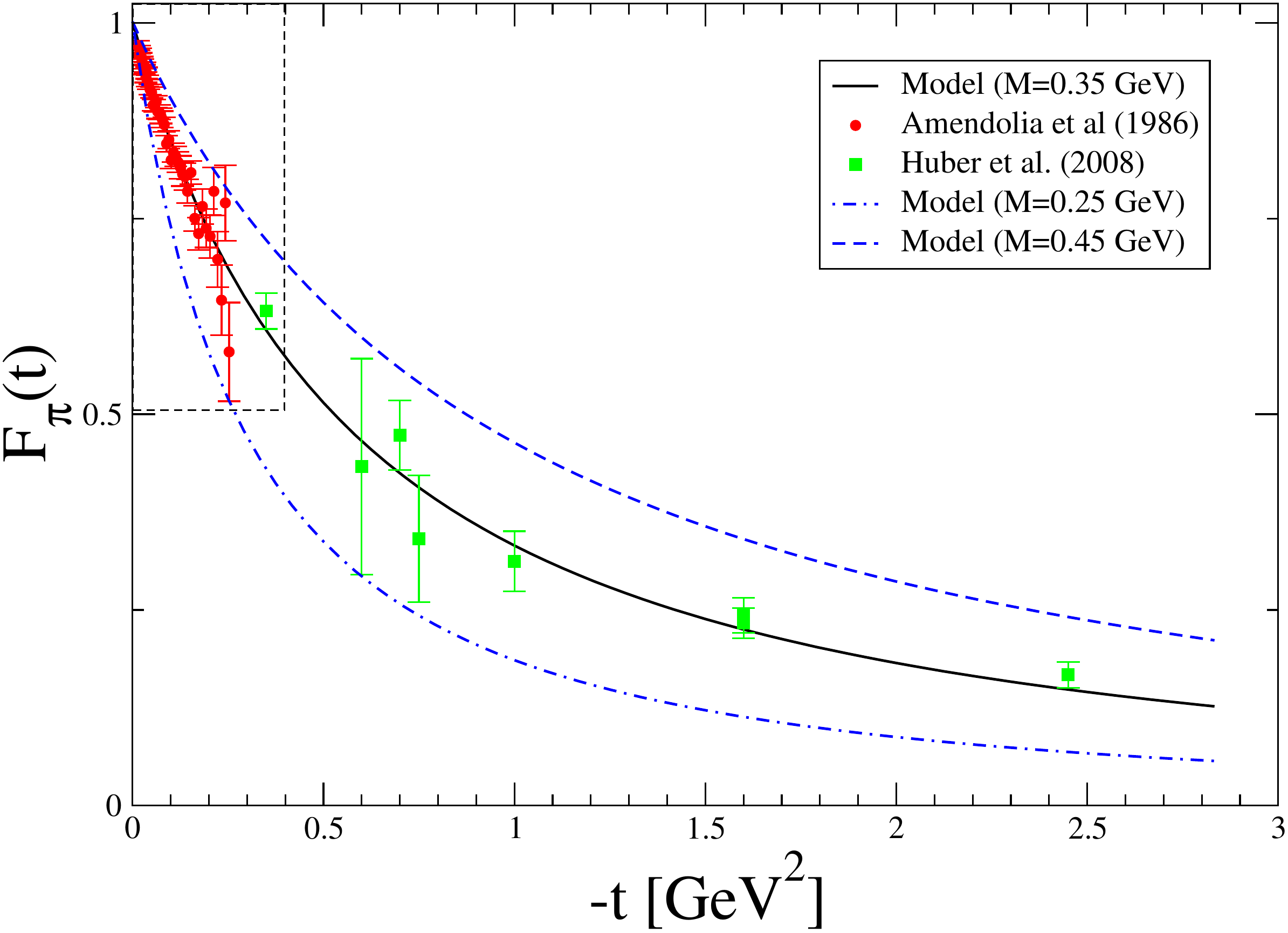} 
  &
  \includegraphics[width= 0.48\textwidth]{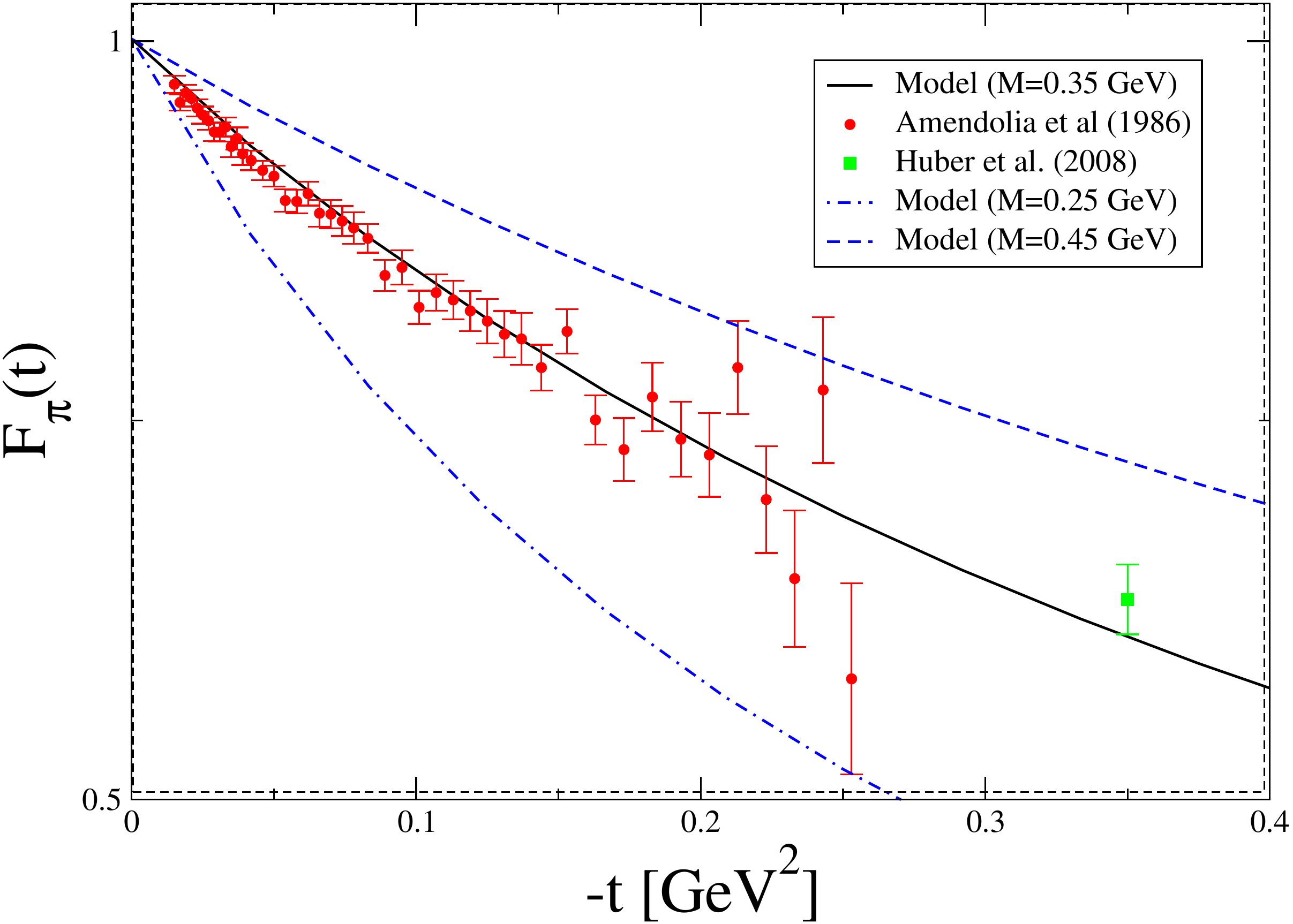} 
  \end{tabular}
  \caption{\small The pion form factor $F_\pi$ computed at $M$ = 0.35~\GeV (solid black line), 0.25~\GeV (dot-dashed blue line) and 0.45~\GeV (dashed blue line), with $\nu = 1$ for the three cases. Experimental data are taken from \refcite{Amendolia:1986wj,Huber:2008id}. The rightmost plot corresponds to a zoom of the dash-delimited area in the leftmost plot allowing to emphasize the constrain provided by the large number of data points in the low-momentum region.}
  \label{fig:FormFactor}
\end{figure}

\reffig{fig:FormFactor} shows a excellent agreement between experimental data and our model for $\nu = 1$ and $M = 0.35~\GeV$. $M$ has a typical constituent quark mass, while $\nu = 1$ corresponds to the case of an asymptotic DA in \refcite{Chang:2013pq}. The model's sensitivity to the quark mass is also displayed by adding the predictions for two more masses roughly in the same ballpark, $M=0.25~\GeV$ and $M=0.45~\GeV$. 
After a simple dimensional analysis with \refeq{eq:pion-electric-charge-radius-na7}, we can see that our model would reach agreement with the NA7 Collaboration value for $M = 339 \pm 3~\MeV$, which is close to our choice of $M = 350~\MeV$. Indeed, we see on \reffig{fig:FormFactor} that our model tends to pass through the upper part of the error bars of the measurements at small $t$, and through the lower part of the error bars at large $t$. 
Presumably a fit varying both $M$ and $\nu$ would permit a perfect agreement to the data at both low $t$, providing a pion charge radius compatible with NA7 data, and large $t$. However such precision studies would not be relevant with our simple model. Let us remind that this algebraic model is the first step towards an implementation of the full numerical solution of the \bs and \dse. The present successful comparison to experimental data is a very encouraging result in our exploratory study.

\subsection{Pion Parton Distribution Function}

The forward limit is another case where our GPD model can be confronted to experimental data since the GPD reduces to the usual PDF:
\begin{equation}
q(x) = H^q(x,0,0). 
\end{equation}
Using \refeq{eq:relation-DD-GPD}, the pion valence PDF ($\beta \geq 0$) can be expressed in terms of DDs:
\begin{equation}
\label{eq:PDF-DD}
q( x ) = 2 \int_{0}^{1-x} \mathrm{d}\alpha \,  F^q( x, \alpha ).
\end{equation}
This yields:
\begin{equation}
q( x ) = \frac{72}{25} \Big( ( 30 - 15 x + 8 x^2 - 2 x^3 )  x^3 \log x + ( 3 + 2 x^2 ) ( 1 - x )^4 \log( 1 - x ) + ( 3 + 15 x + 5 x^2 - 2 x^3 ) x ( 1 - x ) \Big) ,
\label{eq:analytic-PDF-expression}
\end{equation}
which, by construction, has the correct support property, and vanishes at the endpoints. This expression is derived by other means in \refcite{Chang:2014lva}. For large $x$ we observe the following asymptotic behavior:
\begin{equation}
q( x ) \simeq \frac{108}{5} ( 1 - x )^2 \textrm{ when } x \rightarrow 1^-.
\label{eq:asymptotic-analytic-PDF}
\end{equation} 
This  asymptotic $(1-x)^2$ behavior is predicted in the parton model \cite{Ezawa:1974wm,Farrar:1975yb}. Either in perturbative QCD~\cite{Brodsky:1994kg,Ji:2004hz} or within a nonperturbative \ds approach~\cite{Maris:2003vk,Bloch:1999rm,Hecht:2000xa}, a large-$x$ behavior like $(1-x)^{2+\gamma}$ (with an anomalous dimension $\gamma > 0$) is predicted. This is completely consistent with our result in \refeq{eq:asymptotic-analytic-PDF}. The latter is an interesting and very consistent outcome of our simple algebraic model applied to the GPD computation within the \bs and \ds frameworks.

\begin{figure}[!h]
  \begin{center}
	\begin{tabular}{cc}
\includegraphics[width = 0.48\textwidth]{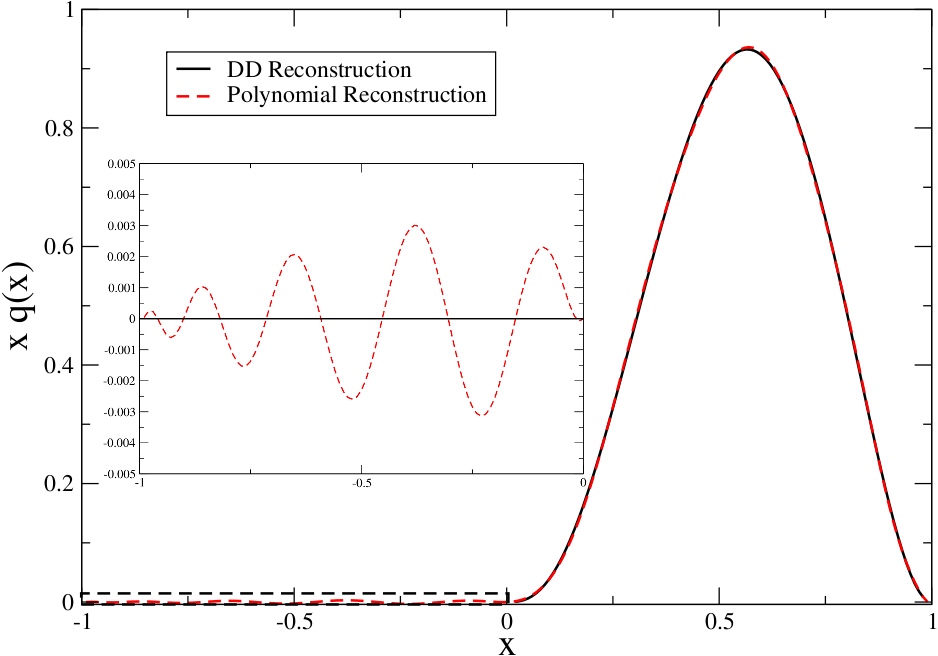} &  
\includegraphics[width = 0.48\textwidth]{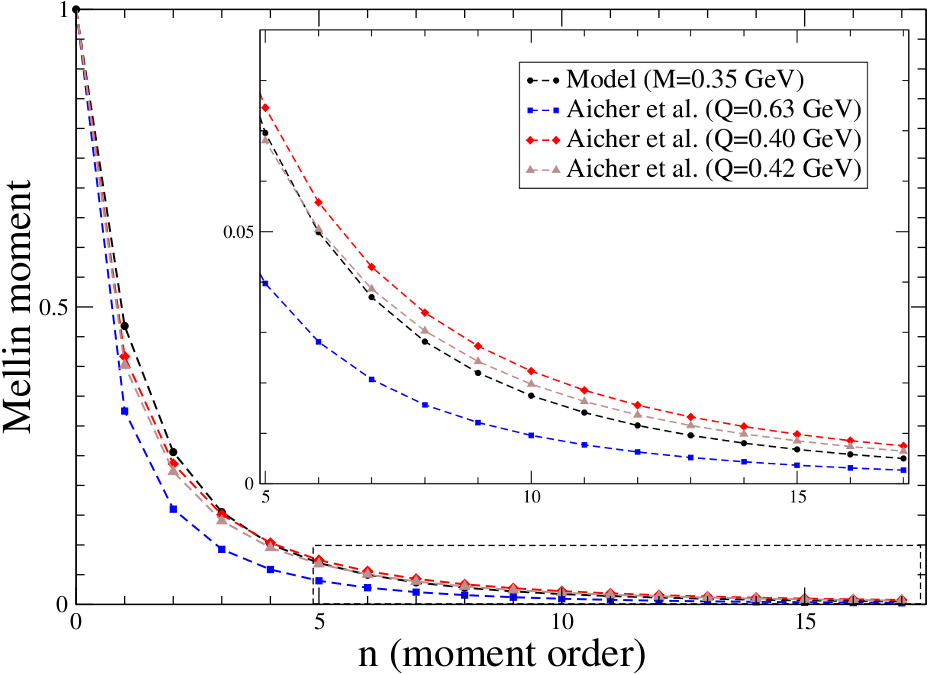}   
	\end{tabular}
	\vspace*{-0.5cm}
  \end{center}
  \caption{\small Left: Pion valence PDF reconstructed from the first Mellin moments evaluated with $\nu=1$ (red dashed line), compared to the exact results 
  obtained from DDs, \refeq{eq:analytic-PDF-expression}. Right: Mellin moments from our model and obtained with the parameterization of Aicher \etal \cite{Aicher:2010cb} run with DGLAP equation down to $Q$ = 0.40~\GeV\xspace and 0.42~\GeV.}
  \label{fig:PDFReconstruction}
\end{figure}

Beyond the intrinsic interest of an expression such as \refeq{eq:analytic-PDF-expression}, we can also start a quantitative discussion of the numerical reconstruction of a PDF from the knowledge of its Mellin moments. Indeed we can get a flavor of the shape of the PDF from the knowledge of the $\simeq 20$ Mellin moments of GPD that we computed with an absolute numerical uncertainty $\simeq 10^{-6}$. Consider for example a Gegenbauer polynomial basis $C_n^{(\alpha)}(x)$:
\begin{equation}
  \label{eq:GegenbauerExpansion}
  H^a(x,0,0) = (1-x^2)^{\alpha-\frac{1}{2}}\sum_{n=0}^{N(\alpha)}d_n^{(\alpha)a}C_n^{(\alpha)}(x) \ ,
\end{equation}
for $a = q$, $(q = u, d)$ or $a = I$, $(I = 0, 1)$.  
The coefficients $d_n^{(\alpha)}$ can be expressed as linear combinations of the first $N(\alpha)$ Mellin moments of the PDF. $\alpha$ and $N(\alpha)$ can be adjusted to obtain a fast-converging series in \refeq{eq:GegenbauerExpansion}. The DD and polynomial reconstructions are compared on the left panel of \reffig{fig:PDFReconstruction}. Indeed, the knowledge of $\simeq 20$ Mellin moments of the PDF already allows a good quantitative reconstruction of the pion valence PDF. However we observe the oscillations typical of an expansion onto a polynomial basis for $x \leq 0$, while the support property is exactly satisfied in the DD approach.

The Mellin moments $\cond{x^m}^q( \xi, t )$ (\ref{eq:MellinMoments}) are dimensionless functions of $\xi$, $t$, $M$ and $\nu$, and thus depend only on $\nu$, $\xi$ and $t/M^2$. Therefore in the forward limit the pion PDF, defined by its Mellin moments, is a function of $x$ and $\nu$ only. In the following we will keep $\nu=1$ and take $M$ fixed at 0.35~\GeV. 

The factorization scale dependence of the GPD has not been set yet since form factors, being observable quantities, do not depend neither on factorization nor renormalization scales. We will now consider that our model is defined at an initial scale $Q_i$ to be determined. At this scale the functional form of the GPD is given by \refeqs{eq:toy-model-quark-propagator}{eq:toy-model-vertex-smearing-function}.

Pion PDFs have been measured by the E615 Collaboration at Fermilab \cite{Conway:1989fs} in a Drell~-~Yan process\footnote{They measured a $\pi^-$ beam interacting with a tungsten target to produce a muon pair, where the longitudinal momentum fraction of the struck quark in the pion is larger than 0.21 and the hard scale, provided by the invariant muon mass, is larger than 4.05~\GeV.} and data have been very recently analysed~\cite{Aicher:2010cb} and shown to yield a reliable parameterization of the pion valence PDF at the low scale $Q_0 = 0.63~\GeV$. Thus, in order to find the (low) definition scale $Q_i$ of our model, we can compute the Mellin moments for this parameterization, run a leading-order DGLAP evolution equation  \cite{Dokshitzer:1977sg,Gribov:1972ri,Altarelli:1977zs} from $Q_0$ downward until we find a scale where the evolved Mellin moments compare well to our model expectation. This study is summarized in the right panel of \reffig{fig:PDFReconstruction}, where we see that $Q_i \simeq 400~\MeV$  allows a good agreement with the PDF extracted from experimental data. Very recently, a refined treatment, so far developped only for the pion PDF in a different context \cite{Chang:2014lva}, improves its agreement with the E615 data.


\section{Conclusion}

GPDs contain unique information about the three-dimensional structure of hadrons and have been triggering a lot of theoretical and experimental activities since the mid-nineties. They are also key-component of the physics cases of Jefferson Lab's upgrade at 12~ \GeV\xspace and of a potential future Electron-Ion Collider. However the computation of GPDs relying on QCD first principles is still an open question, even on the lattice where only few Mellin moments can be evaluated.  This paper is a first step forward in a program aimed at modeling GPDs within the nonperturbative framework of QCD \ds and \bs equations, using well-defined and systematically improvable approximations. 

By definition, GPDs and DDs parameterize matrix elements of nonlocal operators. Here we have followed the Operator Product Expansion approach to expand the pion GPD onto an infinite tower of local operators. Inserted in triangle diagrams, these matrix elements provide systematic expressions for the computation of DDs and of the Mellin moments of pion GPD $H$ at any order. The pion GPD is then recovered from the DDs thanks to a Radon transform. 

The main ingredients for the computations of DDs and GPDs are the full quark propagator and the \bs amplitude. Both should be derived by solving, respectively, the quark gap and the \bs equations. Their implementation for the  computations of DDs and GPDs is left for future work. Here we applied a simple analytical model for both the full quark propagator and the \bs amplitudes. This model, exhibiting most of the analytic features of the realistic \ds and \bs solutions, satisfies several theoretical constraints. In particular the support and polynomiality properties, and the consequences of time-reversal invariance and charge conjugation can be checked analytically. An analytic expression of the pion valence PDF at a low scale is also derived. 

Furthermore, this simple model accounts already well for available experimental data. We successfully compared the zeroth order Mellin moment of the pion GPD to pion form factor measurements for space-like momentum transfer $-t$ between 0 and 2.5~\GeV$^2$. We obtained a remarkable agreement with the data  by adjusting only one parameter encoding a constituent quark mass $M$. 
We also confronted our model PDF to an extraction of the valence PDF from Drell~-~Yan data. The  Mellin moments of the extracted PDF run with DGLAP equations down to a scale $Q_i \simeq 400~\MeV$ are in good agreement with the Mellin moments directly obtained in the model. At last, we have used a basis of Gegenbauer polynomials for a preliminary discussion of the reconstruction of the pion PDF from its Mellin moments, outlining the advantages of our computational strategy involving DDs. Nicely, the computed PDF behaves in the large-$x$ domain in a fully consistent way with previous studies on the pion valence quark PDF using (nonperturbative) numerical solutions of the \dse and perturbative QCD arguments. 

For the scope of the present exploratory study, we retain that our form factor and PDF results are in good agreement with experiment and validate the general functional form of the model. In a forthcoming work, we will present a detailed study of the reconstruction of the pion GPD in connection with Double Distribution models. We will also discuss more closely the aspects related to QCD evolution, allowing in particular comparisons to lattice evaluations of Mellin moments of pion GPDs \cite{Hagler:2009ni}.


\section*{Acknowledgments}

The authors thank A.~Besse, L.~Chang, C.D.~Roberts, P.~Tandy, P.~Fromholz, P.~Kroll, C.~Lorc\'e, J.-Ph.~Lansberg and S.~Wallon for numerous useful and stimulating discussions. They are grateful for the opportunity to participate in the workshop "Many Manifestations of Nonperturbative QCD under the Southern Cross", Ubatuba, S$\tilde{\textrm{a}}$o Paulo, where this work was first presented, and many of its aspects discussed.

This work is partly supported by the Commissariat \`a l'Energie Atomique, the Joint Research Activity "Study of Strongly Interacting Matter" (acronym HadronPhysics3, Grant Agreement n.283286) under the Seventh Framework Programme of the European Community, by the GDR 3034 PH-QCD "Chromodynamique Quantique et Physique des Hadrons", the ANR-12-MONU-0008-01 "PARTONS" and the Spanish ministry Research Project FPA2011-23781.

\appendix 

\section{Analytic expressions for Mellin moments and Double Distributions}
\label{app:A}

We have obtained the following Mellin moments for the isoscalar ($+$) and isovector ($-$) GPDs, as 
explained in \refsec{sec:results}, from \refeq{eq:mellin-moment-direct-diagram-dd}:
\small
\begin{eqnarray}
  \label{eq:Moment}
  \langle x^m \rangle^{I=0,1} & = & \lambda \int \textrm{d}x \, \textrm{d}y \, \textrm{d}u \, \textrm{d}v \, \textrm{d}w \, \textrm{d}z \, \textrm{d}z' \ \left(\frac{M^2}{M'^2}\right)^{2\nu} \delta(1-x-y-u-v-w) x^{\nu -1}y^{\nu-1}\rho(z)\rho(z')
  \nonumber \\
  & & \left[ (g -2\xi~f)^{m}(g+1-2\xi f) \pm (-g -2\xi~f)^{m}(-g-1-2\xi f) \rule[0cm]{0cm}{0.6cm} \right. \nonumber \\
  & & + \frac{1}{2}((-2\xi f + g -1 )(g -2 \xi~f)^{m} \pm (-2\xi f - g +1 )(-g -2 \xi~f)^{m}) \nonumber \\
  & & + \frac{m}{2}\left((g-2\xi f)^{m-1}((g-2\xi f)^2 - \xi^2) \pm (-g-2\xi f)^{m-1}((-g-2\xi f)^2 -\xi^2)\right) \nonumber \\
  & & + \frac{\Gamma(2\nu +1)}{2M'^2\Gamma(2\nu)}(g-2\xi f)^m\left( (g-2\xi f)(tf^2 + P^2(g^2-2g)+\frac{t}{4}+M^2) \right. \nonumber \\
  & & \left.+ tf^2+P^2g^2-\frac{t}{4} + tf\xi+M^2  \right) \pm\frac{\Gamma(2\nu +1)}{2M'^2\Gamma{2\nu}}(-g-2\xi f)^m \left(\rule[0cm]{0cm}{0.50cm} (-g-2\xi f) \right. \nonumber \\
  & & \times (tf^2 + P^2(g^2-2g)+\frac{t}{4}+M^2)\left. \left.- tf^2-P^2g^2+\frac{t}{4} + tf\xi - M^2  \right) \rule[0cm]{0cm}{0.75cm} \right].
\end{eqnarray}
\normalsize
This proves analytically that our algebraic model fulfills the polynomiality property \refeqs{eq:polynomiality_isoscalar_even_power}{eq:polynomiality_isovector_odd_power}. Our result is independent of the parameter $\eta$ appearing in \refeq{eq:TriangleDiagrams} as expected in a Poincar\'e-covariant computation. Then, as explained in the main text, the DDs  can be readily identified from \refeq{eq:Moment} using \refeq{eq:MellinMoments} and \refeq{eq:twist-two-quark-operator-def-DD}: 

\small
\begin{eqnarray}
  \label{eq:DDF}
 F^u(\beta,\alpha,t) & = & \frac{48}{5}\left\{-\frac{18 M^4 t (\beta -1) (\alpha -\beta +1) (\alpha +\beta -1) \left(\left(\alpha ^2-(\beta -1)^2\right) \tanh ^{-1}\left(\frac{2 \beta }{-\alpha ^2+\beta ^2+1}\right)+2 \beta \right)}{\left(4 M^2+t \left((\beta
   -1)^2-\alpha ^2\right)\right)^3} \right.\nonumber \\
& & +\frac{9 M^4 (\alpha -\beta +1) \left(-4 \beta  \left(-\alpha ^2+\beta ^2+1\right)+2 \tanh ^{-1}\left(\frac{2 \beta }{-\alpha ^2+\beta ^2+1}\right)\right)}{4 (\alpha -\beta -1) \left(4 M^2+t \left((\beta
   -1)^2-\alpha ^2\right)\right)^2} \nonumber \\
& & +\frac{9 M^4 (\alpha -\beta +1)\left(\left(\alpha ^4-2 \alpha ^2
   \left(\beta ^2+1\right)+\beta ^2 \left(\beta ^2-2\right)\right) \log \left(\frac{(\alpha -\beta -1) (\alpha +\beta +1)}{\alpha ^2-(\beta -1)^2}\right)\right)}{4 (\alpha -\beta -1) \left(4 M^2+t \left((\beta
   -1)^2-\alpha ^2\right)\right)^2} \nonumber \\
& & +\frac{9 M^4 (\alpha +\beta -1) \left(-4 \beta  \left(-\alpha ^2+\beta ^2+1\right)+2 \tanh ^{-1}\left(\frac{2 \beta }{-\alpha ^2+\beta ^2+1}\right)\right)}{4 (\alpha +\beta +1) \left(4 M^2+t \left((\beta -1)^2-\alpha
   ^2\right)\right)^2} \nonumber \\
& & +\frac{9 M^4 (\alpha +\beta -1)\left(\left(\alpha ^4-2 \alpha ^2
   \left(\beta ^2+1\right)+\beta ^4-2 \beta ^2\right) \log \left(\frac{(\alpha -\beta -1) (\alpha +\beta +1)}{\alpha ^2-(\beta -1)^2}\right)\right)}{4 (\alpha +\beta +1) \left(4 M^2+t \left((\beta -1)^2-\alpha
   ^2\right)\right)^2} \nonumber \\
& & +\left.\frac{9 M^4 \beta  (\alpha -\beta +1)^2 (\alpha +\beta -1)^2 \left(\frac{2 \left(\alpha ^2 \beta -\beta ^3+\beta \right)}{\alpha ^4-2 \alpha ^2 \left(\beta ^2+1\right)+\left(\beta
   ^2-1\right)^2}-\tanh ^{-1}(\alpha -\beta )+\tanh ^{-1}(\alpha +\beta )\right)}{\left(4 M^2+t \left((\beta -1)^2-\alpha ^2\right)\right)^2}\right\}, \nonumber \\
\end{eqnarray}
\begin{eqnarray}
  \label{eq:DDG}
  G^u(\beta,\alpha,t) & = & 
  \frac{48}{5}\left\{-\frac{18 M^4 t \alpha  (\alpha -\beta +1) (\alpha +\beta -1) \left(\left(\alpha ^2-(\beta -1)^2\right) \tanh ^{-1}\left(\frac{2 \beta }{-\alpha ^2+\beta ^2+1}\right)+2 \beta \right)}{\left(4 M^2+t \left((\beta -1)^2-\alpha^2\right)\right)^3}\right. \nonumber \\
  & & -\frac{9 M^4 (\alpha -\beta +1) \left(-4 \beta  \left(-\alpha ^2+\beta ^2+1\right)+2 \tanh ^{-1}\left(\frac{2 \beta }{-\alpha ^2+\beta ^2+1}\right)\right)}{4 (\alpha -\beta -1) \left(4 M^2+t \left((\beta -1)^2-\alpha
   ^2\right)\right)^2} \nonumber \\
  & & -\frac{9 M^4 (\alpha -\beta +1) \left(\left(\alpha ^4-2 \alpha ^2 \left(\beta
   ^2+1\right)+\beta ^2 \left(\beta ^2-2\right)\right) \log \left(\frac{(\alpha -\beta -1) (\alpha +\beta +1)}{\alpha ^2-(\beta -1)^2}\right)\right)}{4 (\alpha -\beta -1) \left(4 M^2+t \left((\beta -1)^2-\alpha
   ^2\right)\right)^2} \nonumber \\
& & +\frac{9 M^4 (\alpha +\beta -1) \left(-4 \beta  \left(-\alpha ^2+\beta ^2+1\right)+2 \tanh ^{-1}\left(\frac{2 \beta }{-\alpha ^2+\beta ^2+1}\right)\right)}{4 (\alpha +\beta +1) \left(4 M^2+t \left((\beta -1)^2-\alpha
   ^2\right)\right)^2} \nonumber \\
& & +\frac{9 M^4 (\alpha +\beta -1)\left(\left(\alpha ^4-2 \alpha ^2 \left(\beta
   ^2+1\right)+\beta ^4-2 \beta ^2\right) \log \left(\frac{(\alpha -\beta -1) (\alpha +\beta +1)}{\alpha ^2-(\beta -1)^2}\right)\right)}{4 (\alpha +\beta +1) \left(4 M^2+t \left((\beta -1)^2-\alpha
   ^2\right)\right)^2} \nonumber \\
& & +\left.\frac{9 M^4 \alpha  (\alpha -\beta +1)^2 (\alpha +\beta -1)^2 \left(\frac{2 \left(\alpha ^2 \beta -\beta ^3+\beta \right)}{\alpha ^4-2 \alpha ^2 \left(\beta ^2+1\right)+\left(\beta
   ^2-1\right)^2}-\tanh ^{-1}(\alpha -\beta )+\tanh ^{-1}(\alpha +\beta )\right)}{\left(4 M^2+t \left((\beta -1)^2-\alpha ^2\right)\right)^2}\right\}. \nonumber \\
\end{eqnarray}
\normalsize
Several comments are in order here. First of all, our model corresponds to a very general case where the DDs do not vanish on the edges of the rhombus, but only at the corners. This is enough to fulfill the support property of GPDs and DDs, the continuity at $x=\xi$ and the vanishing at $| x | = 1$ of the GPD. Indeed the behavior of the GPD at $x = \pm 1$ or $x = \pm \xi$ is related to the analytic properties of the DDs at the vertices of the rhombus $\Omega$. More precisely\footnote{We will make repeated use of the following relation:
\begin{displaymath}
\textrm{When } (b-a) \rightarrow 0 \quad \theta( a \leq z \leq b ) \simeq ( b - a ) \delta\left( z - \frac{a+b}{2} \right).
\end{displaymath}%
} we see that:
\begin{equation}
\label{eq:DD-vertex-endpoint-x}
H^q( x, \xi ) \simeq ( 1 - x ) \frac{2}{1 - \xi^2} \big( F^q( 1, 0 ) + \xi G^q( 1, 0 ) \big) \textrm{ when } x \rightarrow 1^-,
\end{equation}
and:
\begin{eqnarray}
\label{eq:DD-vertex-xi-x}
H^q( x, \xi ) - H^q( \xi, \xi )
& \simeq & 
\frac{x - \xi}{\xi} \left( \int_{\frac{x-\xi}{1-\xi}}^{\frac{x+\xi}{1+\xi}} \, \mathrm{d}\beta \left[ \frac{1}{\xi} \partial_\alpha F^q\left( \beta, \frac{\xi - \beta}{\xi} \right) + \partial_\alpha G^q\left( \beta, \frac{\xi - \beta}{\xi} \right) \right] \right. \nonumber \\
& & \left. + \frac{-2\xi}{1 - \xi^2} \big( F^q( 0, 1 ) + \xi G^q( 0, 1 ) \big) \right) \textrm{ when } x \rightarrow \xi,
\end{eqnarray}
with similar relations for $x$ close to $-1$ or $-\xi$. The vanishing of the pion GPD $H^q$ at $x = \pm 1$ has been established in perturbative QCD \cite{Yuan:2003fs} with a polynomial fall-off:
\begin{equation}
\label{eq:pion-gpd-x-endpoint}
H^q( x, \xi )  \simeq \frac{( 1 - x )^2}{1 - \xi^2} \textrm{ when } x \rightarrow 1^-.
\end{equation}
We see from \refeq{eq:DD-vertex-endpoint-x} that the pion GPD $H^q$ vanishes at $x = \pm 1$ as soon as the DDs $F^q$ and $G^q$ are finite or not too singular at $( \beta, \alpha ) = ( \pm 1, 0 )$. The perturbative behavior (\ref{eq:pion-gpd-x-endpoint}) is matched when the DDs vanish fast enough. Similarly the continuity of the GPD $H^q$ near $x = \pm \xi$, necessary to the factorization of the DVCS amplitude \cite{Radyushkin:1997ki}, requires that the DDs $F^q$ and $G^q$ are not too singular at $( \beta, \alpha ) = ( \pm 0, 1 )$. This last observation has already been made in \refcite{Radyushkin:1997ki,Radyushkin:1998bz,Radyushkin:2000uy}.

\bibliographystyle{h-physrev5}
\bibliography{Bibliography}

\begin{thebibliography}{10}

\bibitem{Mueller:1998fv}
D.~Mueller, D.~Robaschik, B.~Geyer, F.~Dittes, and J.~Ho\v{r}e\v{j}si,
\newblock Fortsch.Phys. {\bf 42}, 101 (1994), arXiv:hep-ph/9812448.

\bibitem{Ji:1996nm}
X.-D. Ji,
\newblock Phys.Rev. {\bf D55}, 7114 (1997), arXiv:hep-ph/9609381.

\bibitem{Radyushkin:1997ki}
A.~Radyushkin,
\newblock Phys.Rev. {\bf D56}, 5524 (1997), arXiv:hep-ph/9704207.

\bibitem{Ji:1998pc}
X.-D. Ji,
\newblock J.Phys. {\bf G24}, 1181 (1998), arXiv:hep-ph/9807358.

\bibitem{Goeke:2001tz}
K.~Goeke, M.~V. Polyakov, and M.~Vanderhaeghen,
\newblock Prog.Part.Nucl.Phys. {\bf 47}, 401 (2001), arXiv:hep-ph/0106012.

\bibitem{Diehl:2003ny}
M.~Diehl,
\newblock Phys.Rept. {\bf 388}, 41 (2003), arXiv:hep-ph/0307382.

\bibitem{Belitsky:2005qn}
A.~Belitsky and A.~Radyushkin,
\newblock Phys.Rept. {\bf 418}, 1 (2005), arXiv:hep-ph/0504030.

\bibitem{Boffi:2007yc}
S.~Boffi and B.~Pasquini,
\newblock Riv.Nuovo Cim. {\bf 30}, 387 (2007), arXiv:0711.2625.

\bibitem{Guidal:2013rya}
M.~Guidal, H.~Moutarde, and M.~Vanderhaeghen,
\newblock Rept.Prog.Phys. {\bf 76}, 066202 (2013), arXiv:1303.6600.

\bibitem{Radyushkin:1998es}
A.~Radyushkin,
\newblock Phys.Rev. {\bf D59}, 014030 (1999), arXiv:hep-ph/9805342.

\bibitem{Radyushkin:1998bz}
A.~Radyushkin,
\newblock Phys.Lett. {\bf B449}, 81 (1999), arXiv:hep-ph/9810466.

\bibitem{Teryaev:2001qm}
O.~Teryaev,
\newblock Phys.Lett. {\bf B510}, 125 (2001), arXiv:hep-ph/0102303.

\bibitem{Broniowski:2007si}
W.~Broniowski, E.~Ruiz~Arriola, and K.~Golec-Biernat,
\newblock Phys.Rev. {\bf D77}, 034023 (2008), arXiv:0712.1012.

\bibitem{Frederico:2009fk}
T.~Frederico, E.~Pace, B.~Pasquini, and G.~Salme,
\newblock Phys.Rev. {\bf D80}, 054021 (2009), arXiv:0907.5566.

\bibitem{Goldstein:2010gu}
G.~R. Goldstein, J.~O.~G. Hernandez, and S.~Liuti,
\newblock Phys.Rev. {\bf D84}, 034007 (2011), arXiv:1012.3776.

\bibitem{Mezrag:2013mya}
C.~Mezrag, H.~Moutarde, and F.~Sabati\'e,
\newblock Phys.Rev. {\bf D88}, 014001 (2013), arXiv:1304.7645.

\bibitem{Kumericki:2008di}
K.~Kumericki, D.~Mueller, and K.~Passek-Kumericki,
\newblock Eur.Phys.J. {\bf C58}, 193 (2008), arXiv:0805.0152.

\bibitem{Roberts:1994dr}
C.~D. Roberts and A.~G. Williams,
\newblock Prog.Part.Nucl.Phys. {\bf 33}, 477 (1994), arXiv:hep-ph/9403224.

\bibitem{Alkofer:2000wg}
R.~Alkofer and L.~von Smekal,
\newblock Phys.Rept. {\bf 353}, 281 (2001), arXiv:hep-ph/0007355.

\bibitem{Maris:2003vk}
P.~Maris and C.~D. Roberts,
\newblock Int.J.Mod.Phys. {\bf E12}, 297 (2003), arXiv:nucl-th/0301049.

\bibitem{Bashir:2012fs}
A.~Bashir {\em et~al.},
\newblock Commun.Theor.Phys. {\bf 58}, 79 (2012), arXiv:1201.3366.

\bibitem{Roberts:2012sv}
C.~D. Roberts,
\newblock (2012), arXiv:1203.5341.

\bibitem{Dyson:1949ha}
F.~Dyson,
\newblock Phys.Rev. {\bf 75}, 1736 (1949).

\bibitem{Schwinger:1951ex}
J.~S. Schwinger,
\newblock Proc.Nat.Acad.Sci. {\bf 37}, 452 (1951).

\bibitem{Schwinger:1951hq}
J.~S. Schwinger,
\newblock Proc.Nat.Acad.Sci. {\bf 37}, 455 (1951).

\bibitem{Salpeter:1951sz}
E.~Salpeter and H.~Bethe,
\newblock Phys.Rev. {\bf 84}, 1232 (1951).

\bibitem{GellMann:1951rw}
M.~Gell-Mann and F.~Low,
\newblock Phys.Rev. {\bf 84}, 350 (1951).

\bibitem{Schwinger:1953tb}
J.~S. Schwinger,
\newblock Phys.Rev. {\bf 91}, 713 (1953).

\bibitem{Amrath:2008vx}
D.~Amrath, M.~Diehl, and J.-P. Lansberg,
\newblock Eur.Phys.J. {\bf C58}, 179 (2008), arXiv:0807.4474.

\bibitem{Polyakov:1999gs}
M.~V. Polyakov and C.~Weiss,
\newblock Phys.Rev. {\bf D60}, 114017 (1999), arXiv:hep-ph/9902451.

\bibitem{Anikin:1999pf}
I.~Anikin, A.~Dorokhov, A.~Maximov, L.~Tomio, and V.~Vento,
\newblock (1999), arXiv:hep-ph/9905332.

\bibitem{Broniowski:2003rp}
W.~Broniowski and E.~Ruiz~Arriola,
\newblock Phys.Lett. {\bf B574}, 57 (2003), arXiv:hep-ph/0307198.

\bibitem{RuizArriola:2002wr}
E.~Ruiz~Arriola,
\newblock Acta Phys.Polon. {\bf B33}, 4443 (2002), arXiv:hep-ph/0210007.

\bibitem{Christov:1995vm}
C.~Christov {\em et~al.},
\newblock Prog.Part.Nucl.Phys. {\bf 37}, 91 (1996), arXiv:hep-ph/9604441.

\bibitem{Choi:2001fc}
H.-M. Choi, C.-R. Ji, and L.~Kisslinger,
\newblock Phys.Rev. {\bf D64}, 093006 (2001), arXiv:hep-ph/0104117.

\bibitem{Choi:2002ic}
H.-M. Choi, C.-R. Ji, and L.~Kisslinger,
\newblock Phys.Rev. {\bf D66}, 053011 (2002), arXiv:hep-ph/0204321.

\bibitem{Mukherjee:2002gb}
A.~Mukherjee, I.~Musatov, H.~Pauli, and A.~Radyushkin,
\newblock Phys.Rev. {\bf D67}, 073014 (2003), arXiv:hep-ph/0205315.

\bibitem{Ji:2006ea}
C.-R. Ji, Y.~Mishchenko, and A.~Radyushkin,
\newblock Phys.Rev. {\bf D73}, 114013 (2006), arXiv:hep-ph/0603198.

\bibitem{Tiburzi:2002tq}
B.~Tiburzi and G.~Miller,
\newblock Phys.Rev. {\bf D67}, 113004 (2003), arXiv:hep-ph/0212238.

\bibitem{Theussl:2002xp}
L.~Theussl, S.~Noguera, and V.~Vento,
\newblock Eur.Phys.J. {\bf A20}, 483 (2004), arXiv:nucl-th/0211036.

\bibitem{Bissey:2002yr}
F.~Bissey {\em et~al.},
\newblock Phys.Lett. {\bf B547}, 210 (2002), arXiv:hep-ph/0207107.

\bibitem{Bissey:2003yr}
F.~Bissey, J.~Cudell, J.~Cugnon, J.~Lansberg, and P.~Stassart,
\newblock Phys.Lett. {\bf B587}, 189 (2004), arXiv:hep-ph/0310184.

\bibitem{VanDyck:2007jt}
A.~Van~Dyck, T.~Van~Cauteren, and J.~Ryckebusch,
\newblock Phys.Lett. {\bf B662}, 413 (2008), arXiv:0710.2271.

\bibitem{Musatov:1999xp}
I.~Musatov and A.~Radyushkin,
\newblock Phys.Rev. {\bf D61}, 074027 (2000), arXiv:hep-ph/9905376.

\bibitem{Bakulev:2000eb}
A.~P. Bakulev, R.~Ruskov, K.~Goeke, and N.~Stefanis,
\newblock Phys.Rev. {\bf D62}, 054018 (2000), arXiv:hep-ph/0004111.

\bibitem{Vogt:2001if}
C.~Vogt,
\newblock Phys.Rev. {\bf D64}, 057501 (2001), arXiv:hep-ph/0101059.

\bibitem{Hoodbhoy:2003uu}
P.~Hoodbhoy, X.-d. Ji, and F.~Yuan,
\newblock Phys.Rev.Lett. {\bf 92}, 012003 (2004), arXiv:hep-ph/0309085.

\bibitem{Diehl:2005rn}
M.~Diehl, A.~Manashov, and A.~Schafer,
\newblock Phys.Lett. {\bf B622}, 69 (2005), arXiv:hep-ph/0505269.

\bibitem{Chang:2013pq}
L.~Chang {\em et~al.},
\newblock Phys.Rev.Lett. {\bf 110}, 132001 (2013), arXiv:1301.0324.

\bibitem{Moutarde:2013qs}
H.~Moutarde, B.~Pire, F.~Sabatie, L.~Szymanowski, and J.~Wagner,
\newblock Phys.Rev. {\bf D87}, 054029 (2013), arXiv:1301.3819.

\bibitem{Bashir:2013zha}
A.~Bashir, A.~Raya, and J.~Rodriguez-Quintero,
\newblock Phys.Rev. {\bf D88}, 054003 (2013), arXiv:1302.5829.

\bibitem{Conway:1989fs}
J.~Conway {\em et~al.},
\newblock Phys.Rev. {\bf D39}, 92 (1989).

\bibitem{Bordalo:1987cr}
NA10 Collaboration, P.~Bordalo {\em et~al.},
\newblock Phys.Lett. {\bf B193}, 373 (1987).

\bibitem{Betev:1985pf}
NA10 Collaboration, B.~Betev {\em et~al.},
\newblock Z.Phys. {\bf C28}, 9 (1985).

\bibitem{Mandelstam:1955sd}
S.~Mandelstam,
\newblock Proc.Roy.Soc.Lond. {\bf A233}, 248 (1955).

\bibitem{Lurie:1965}
D.~Luri\'e, A.~J. Macfarlane, and Y.~Takahashi,
\newblock Phys. Rev. {\bf 140}, B1091 (1965).

\bibitem{Nishijima:1967}
K.~Nishijima and A.~H. Singh,
\newblock Phys. Rev. {\bf 162}, 1740 (1967).

\bibitem{Amendolia:1986wj}
NA7 Collaboration, S.~Amendolia {\em et~al.},
\newblock Nucl.Phys. {\bf B277}, 168 (1986).

\bibitem{Huber:2008id}
Jefferson Lab, G.~Huber {\em et~al.},
\newblock Phys.Rev. {\bf C78}, 045203 (2008), arXiv:0809.3052.

\bibitem{Chang:2014lva}
L.~Chang {\em et~al.},
\newblock (2014), arXiv:1406.5450.

\bibitem{Ezawa:1974wm}
Z.~Ezawa,
\newblock Nuovo Cim. {\bf A23}, 271 (1974).

\bibitem{Farrar:1975yb}
G.~R. Farrar and D.~R. Jackson,
\newblock Phys.Rev.Lett. {\bf 35}, 1416 (1975).

\bibitem{Brodsky:1994kg}
S.~J. Brodsky, M.~Burkardt, and I.~Schmidt,
\newblock Nucl.Phys. {\bf B441}, 197 (1995), arXiv:hep-ph/9401328.

\bibitem{Ji:2004hz}
X.-d. Ji, J.-P. Ma, and F.~Yuan,
\newblock Phys.Lett. {\bf B610}, 247 (2005), arXiv:hep-ph/0411382.

\bibitem{Bloch:1999rm}
J.~C. Bloch, C.~D. Roberts, and S.~Schmidt,
\newblock Phys.Rev. {\bf C61}, 065207 (2000), arXiv:nucl-th/9911068.

\bibitem{Hecht:2000xa}
M.~Hecht, C.~D. Roberts, and S.~Schmidt,
\newblock Phys.Rev. {\bf C63}, 025213 (2001), arXiv:nucl-th/0008049.

\bibitem{Aicher:2010cb}
M.~Aicher, A.~Schafer, and W.~Vogelsang,
\newblock Phys.Rev.Lett. {\bf 105}, 252003 (2010), arXiv:1009.2481.

\bibitem{Dokshitzer:1977sg}
Y.~L. Dokshitzer,
\newblock Sov.Phys.JETP {\bf 46}, 641 (1977).

\bibitem{Gribov:1972ri}
V.~Gribov and L.~Lipatov,
\newblock Sov.J.Nucl.Phys. {\bf 15}, 438 (1972).

\bibitem{Altarelli:1977zs}
G.~Altarelli and G.~Parisi,
\newblock Nucl.Phys. {\bf B126}, 298 (1977).

\bibitem{Hagler:2009ni}
P.~Hagler,
\newblock Phys.Rept. {\bf 490}, 49 (2010), arXiv:0912.5483.

\bibitem{Yuan:2003fs}
F.~Yuan,
\newblock Phys.Rev. {\bf D69}, 051501 (2004), arXiv:hep-ph/0311288.

\bibitem{Radyushkin:2000uy}
A.~Radyushkin,
\newblock (2000), arXiv:hep-ph/0101225.

\end{thebibliography}

\end{document}